\documentclass[conference]{IEEEtran}
\IEEEoverridecommandlockouts
\usepackage{cite}
\usepackage{amsmath,amssymb,amsfonts}
\usepackage{graphicx}
\usepackage{textcomp}
\usepackage{xcolor}
\usepackage{cleveref}

\usepackage{bm}
\usepackage[linesnumbered,boxed,vlined,ruled,noend]{algorithm2e}
\usepackage{multirow}
\usepackage{balance}
\usepackage{tikz}
\usepackage{url}
\usepackage{xspace}
\newcommand{\sys}{\textsc{Zoomer}\xspace}
\newcommand*{\circled}[1]{\lower.7ex\hbox{\tikz\draw (0pt, 0pt)circle (.5em) node {\makebox[1em][c]{\small #1}};}}
\usepackage{xspace}
\usepackage{multirow}
\usepackage{bm}
\usepackage{bbm}
\usepackage{color, soul}
\usepackage{subfigure}
\usepackage{graphicx}
\usepackage{listings}
\usepackage{xcolor}
\usepackage{makecell}
\usepackage{color, soul}

\usepackage{bm}
\usepackage{booktabs}
\usepackage{wrapfig}
\usepackage{enumitem}
\usepackage{CJKutf8}
\usepackage{textcomp}

\newcommand{\para}[1]{{\vspace{2pt} \bf \noindent #1 \hspace{1pt}}}

\usepackage[linesnumbered,vlined,ruled,noend]{algorithm2e}

\def\BibTeX{{\rm B\kern-.05em{\sc i\kern-.025em b}\kern-.08em
    T\kern-.1667em\lower.7ex\hbox{E}\kern-.125emX}}
\begin{document}

\title{\sys: Boosting Retrieval on Web-scale Graphs by Regions of Interest
}




\author{

\IEEEauthorblockN{
${^\flat  {^\dag}}$Yuezihan Jiang{$^\star$}~~~~~${^\sharp {^\dag}}$Yu Cheng{$^\star$}~~~~~${^\sharp {^\dag}}$Hanyu Zhao~~~~~${^\sharp}$Wentao Zhang~~~~~${^\sharp}$Xupeng Miao~~~~~\\
${^\dag}$Yu He~~~~~${^\dag}$Liang Wang~~~~~${^\sharp {^\ddag}}$Zhi  Yang$^*$~~~~~${^\sharp {^\ddag}}$Bin Cui$^*$\thanks{$^\star$ Co-first authors.}\thanks{ * Corresponding authors.}}

\IEEEauthorblockA{\textit{${^\flat}$School of Software and Microelectronics, Peking University} \\
\textit{${^\sharp}$School of CS \& Key Laboratory of High Confidence Software Technologies, Peking University} \\
\textit{${^\ddag}$Center for Data Science, Peking University \& National Engineering Laboratory for Big Data Analysis and Applications} \\
\textit{$^{\dag}$Alibaba Group} \\
}
\textit{${^\sharp}$ \{jiangyuezihan, yu.cheng, zhaohanyu, wentao.zhang, xupeng.miao, yangzhi, bin.cui\}@pku.edu.cn,}\\
\textit{
${^\dag}$\{herve.hy, liangbo.wl\}@alibaba-inc.com}\\
}
\maketitle

\begin{abstract}
We introduce \sys, a system deployed at Taobao, the largest e-commerce platform in China, for training and serving GNN-based recommendations over web-scale graphs. \sys is designed for tackling two challenges presented by the massive user data at Taobao: low training/serving efficiency due to the huge scale of the graphs, and low recommendation quality due to the information overload which distracts the recommendation model from specific user intentions.
\sys achieves this by introducing a key concept, \textit{Region of Interests} (ROI) in GNNs for recommendations, i.e., a neighborhood region in the graph with significant relevance to a strong user intention. \sys narrows the focus from the whole graph and ``zooms in'' on the more relevant ROIs, thereby reducing the training/serving cost and mitigating the information overload at the same time. With carefully designed mechanisms, \sys identifies the interest expressed by each recommendation request, constructs an ROI subgraph by sampling with respect to the interest, and guides the GNN to reweigh different parts of the ROI towards the interest by a multi-level attention module. Deployed as a large-scale distributed system, \sys supports graphs with billions of nodes for training and thousands of requests per second for serving. \sys achieves up to 14x speedup when downsizing sampling scales with comparable (even better) AUC performance than baseline methods. Besides, both the offline evaluation and online A/B test demonstrate the effectiveness of \sys.
\end{abstract}


\section{Introduction}
Graph Neural Networks (GNNs) have recently gained much momentum for their superiority in Recommendation Systems (RS)  \cite{wu2020graph, liu2020basket, cao2022geometric, yang2019aligraph}, since interaction data in RS could be naturally regarded as a graph between user nodes and item nodes with edges representing interactions. Due to the state-of-the-art performance in many graph-based tasks like link prediction, which is suitable for converting into the Click-Through Rate (CTR) prediction task in RS, GNN-based recommendations has been an active research topic in the information retrieval community and search engine industry as the next-generation search technology  \cite{wang2019deep, ying2018graph, yang2020multisage}. In this work, we consider the problem of developing and deploying GNN-based recommendation systems in a real industry application, Taobao, the largest e-commerce platform in China. 
Different from graphs in academy research, 
graph data at modern web companies poses extra challenges to GNN-based recommendation:

\para{Huge Scale.} Rich user behavior data is important to click-through prediction tasks\cite{pi2020search}. However, under the guidance of long-term user behaviors logs, graphs often contain billions of nodes and
trillions of edges, which brings significant engineering obstacles to scaling GNN-based systems.
For example, the Facebook graph includes over two billion
user nodes and over a trillion edges representing friendships, likes, posts, and other connections~\cite{ching2015one}. At Pinterest, the user-to-item graph includes over 2 billion entities and over 17 billion edges~\cite{ying2018graph}. In Taobao's real production environment, the graph of users and products also consists of one billion nodes and ten billion edges. 
Graphs with these scales pose great challenges because most GNNs adopt the expensive recursive neighborhood expansion through the graph, leading to unacceptable training costs on web-scale networks~\cite{chen2021towards, zhang2021graph, zhang2021node}. 



\para{Information Overload.} 
Another challenge in the real production environment is information overload, which means the excess of available information to make an effective decision. 
For example, at Taobao, as long-term interests have been proved helpful for user modeling~\cite{pi2020search}, RS takes historical logs from a longer span as model input. Besides, both users and products are represented by  extensive features in a fine-grained way. Moreover, each node in a Taobao graph generally connects to extensive nodes and has multiple types of relations (e.g., clicks and co-occurrence). However, search platform is the most deterministic scenario in RS, because users in search platforms have strongly specific intentions for certain kinds of products. The information-rich nature of the Taobao environment and the direct demands for products in search platforms can easily become traps for information overload, requiring effectively filtering useless data.

To address these challenges, we advocate a new concept of \emph{Regions of Interest (ROI)} for GNN-based recommendation, which is an area containing more important information. 
To help illustrate this idea, consider the example of Taobao, where users can search products by posing queries, the system returns a set of items related to the posed queries, and users provide feedback by browses and clicks, thus creating a heterogeneous user-query-item graph. 
In such a graph, we define an ROI as a relevant neighborhood region with significant relevance to the given focal points of interest. 
Here, we define focal points of interest as nodes of the user and the corresponding time-sensitive query requesting for recommendation and their neighbors which could represent the current search intention.



\begin{figure}[tpb!]
    \centering
    \includegraphics[width=0.7\linewidth]{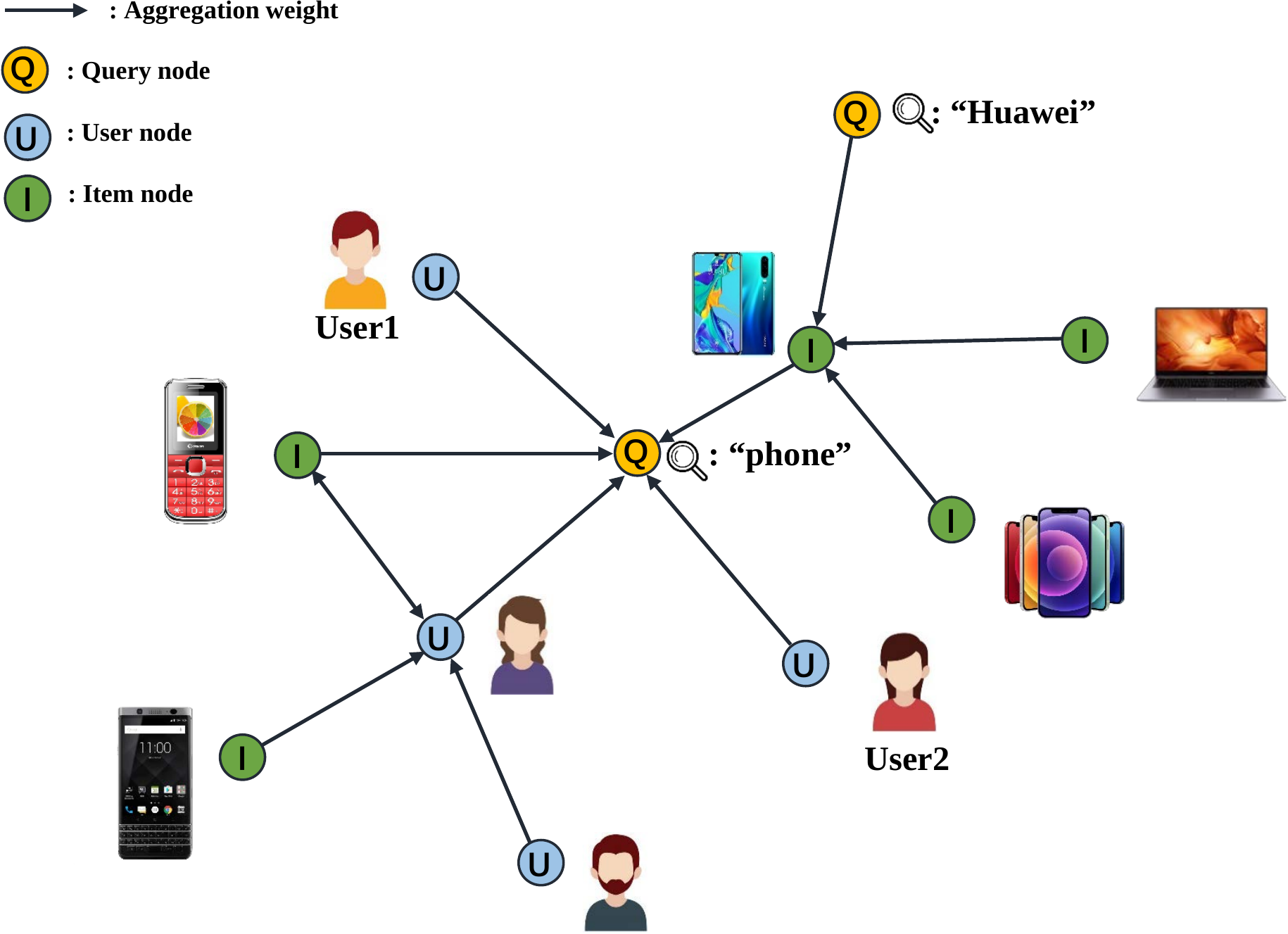}
    \caption{General aggregations of classical GNNs.}
    \label{fig:classical}
\end{figure}

\begin{figure}[tpb!]
    \centering
    \includegraphics[width=0.7\linewidth]{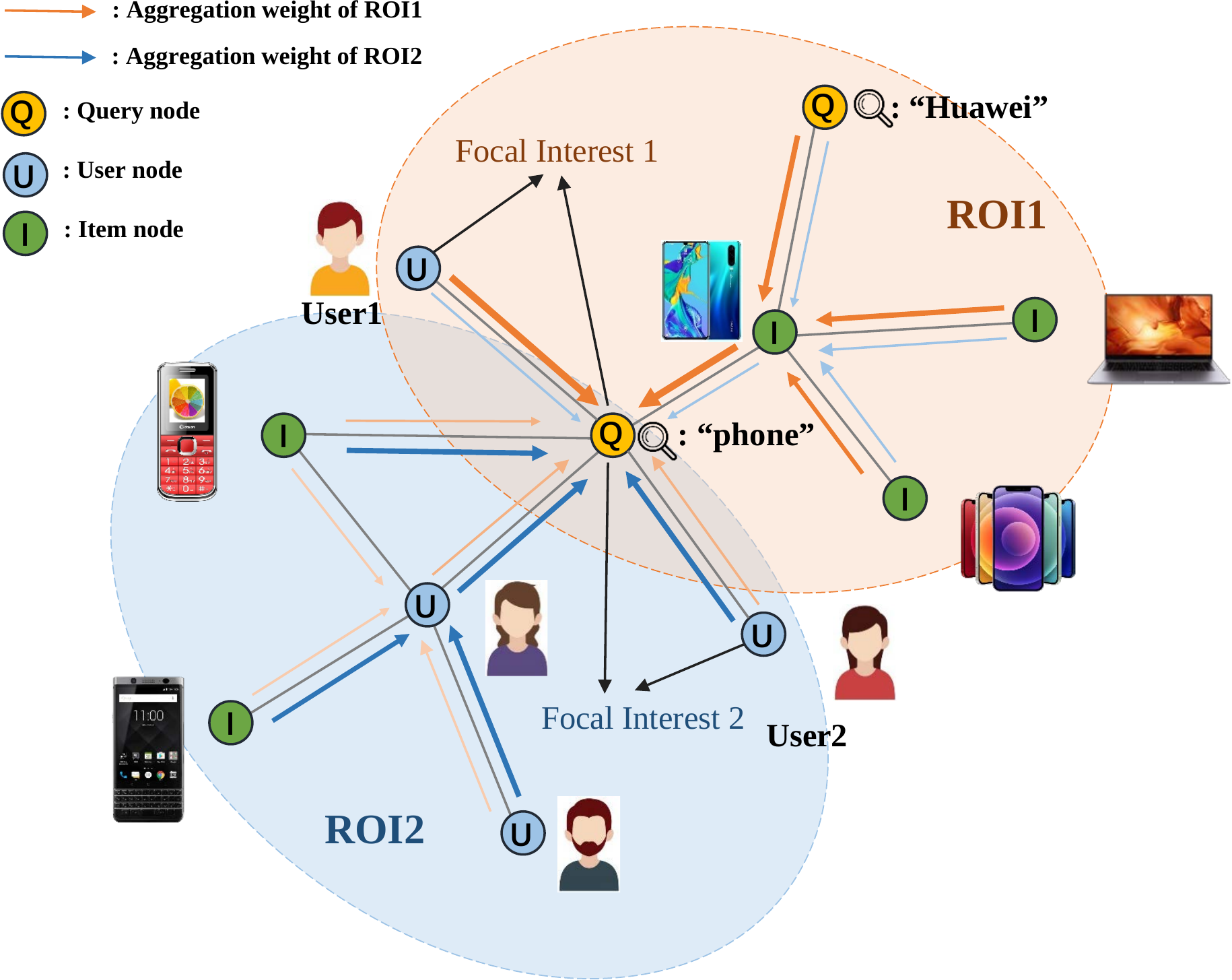}
    \caption{ROI-based aggregation. The thicker the arrows, the greater the edge weights. }
    \label{fig:ours}
\end{figure}

Focusing attention on the ROI region empowers GNNs to handle long-term user interests and reduce both the cost of neighborhood expansion and information overload. Fig.~\ref{fig:classical} illustrates the aggregation process of standard GNNs, which mixes all nodes in the neighborhood region into a single fixed embedding. For example, the embedding of the ``phone'' query node for both User1 and User2 is fixed due to the static aggregation--nodes in the $K$-hop neighborhood region are aggregated through pooling given a $K$-layer GNN. The key limitation of existing GNNs is that they do not sufficiently distinguish intention-related node interactions related to the dynamic search purpose. 
By contrast, with the concept of ROI, we can dynamically compute multiple embeddings that capture the diverse intentions and characteristics implied by different focal points during graph convolution for each ego node. Fig.~\ref{fig:ours} illustrates that for the ego query node ``phone'',  we complement User1 and User2 separately as the focal of interests, and obtain the corresponding multiple embeddings for the query node by identifying different focal-related sub-graphs and aggregating its neighborhood region in different ways, e.g., putting more attention on the neighborhood regions more relevant to the focal points. 

To effectively leverage ROIs, we propose a powerful Recommendation System \sys, which flexibly and adaptively computes suitable embeddings for the ego nodes based on the focal points of interest. Deployed in a distributed architecture, \sys performs efficient, localized convolutions by sampling the neighborhoods with significant relevance with users' focal interests to construct ROI. By constructing localized ROI out of user intention and their personalized characteristics, we mitigate the data overload and avoid the operations on the entire graph during the GNN model training process. Therefore, \sys is able to support web-scale Taobao graphs with billion-scale nodes and tens of billion-scale edges at an acceptable cost.
To address the information overload problem, we extend GNN by incorporating a focal-oriented filter operation into its neighborhood aggregation process, where we perform a fine-grained ROI identification and dynamically compute multiple embeddings for each target node under the intentions and user-specific characteristics implied by different focal interest points. 
Particularly, we design a novel attention mechanism with flexible node-level feature embedding projection, edge-level neighbor reweighing, and semantic-level relation combination during graph convolutions.

We implement \sys in a distributed framework and deploy it for both offline training and online serving. The offline training module
allows scalable model training on massive networks with
millions to billions of nodes.
For the online module, \sys could support personal items retrieval for thousands of search requests per second. 

Our contributions can be summarized as follows: (1)~\emph{Concepts and mechanisms.} To the best of our knowledge, we are the first to propose the definition of focal interests and ROI for GNN-based recommendations. Based on focal interests, we present an effective sampling strategy for constructing ROIs from users' long-term interests and a multi-level attention mechanism to perform both localized and intention-aware convolutions.
(2)~\emph{Web-scale system and implementation.} We introduce our hands-on experience of designing and implementing an ROI-based industrial system, \sys, to support web-scale recommendations and deploy it in the production environment at Taobao. 
(3)~\emph{Performance and efficiency.} \sys could effectively speed up training on long-term user interests with comparable (or even better) performance. Besides, \sys supports millisecond inference response for online serving. Both the offline evaluation and online A/B test demonstrate the effectiveness of \sys.



\section{Problem Setup}\label{sec:graph}

\begin{figure}
	\centering
	\includegraphics[width=0.9\columnwidth]{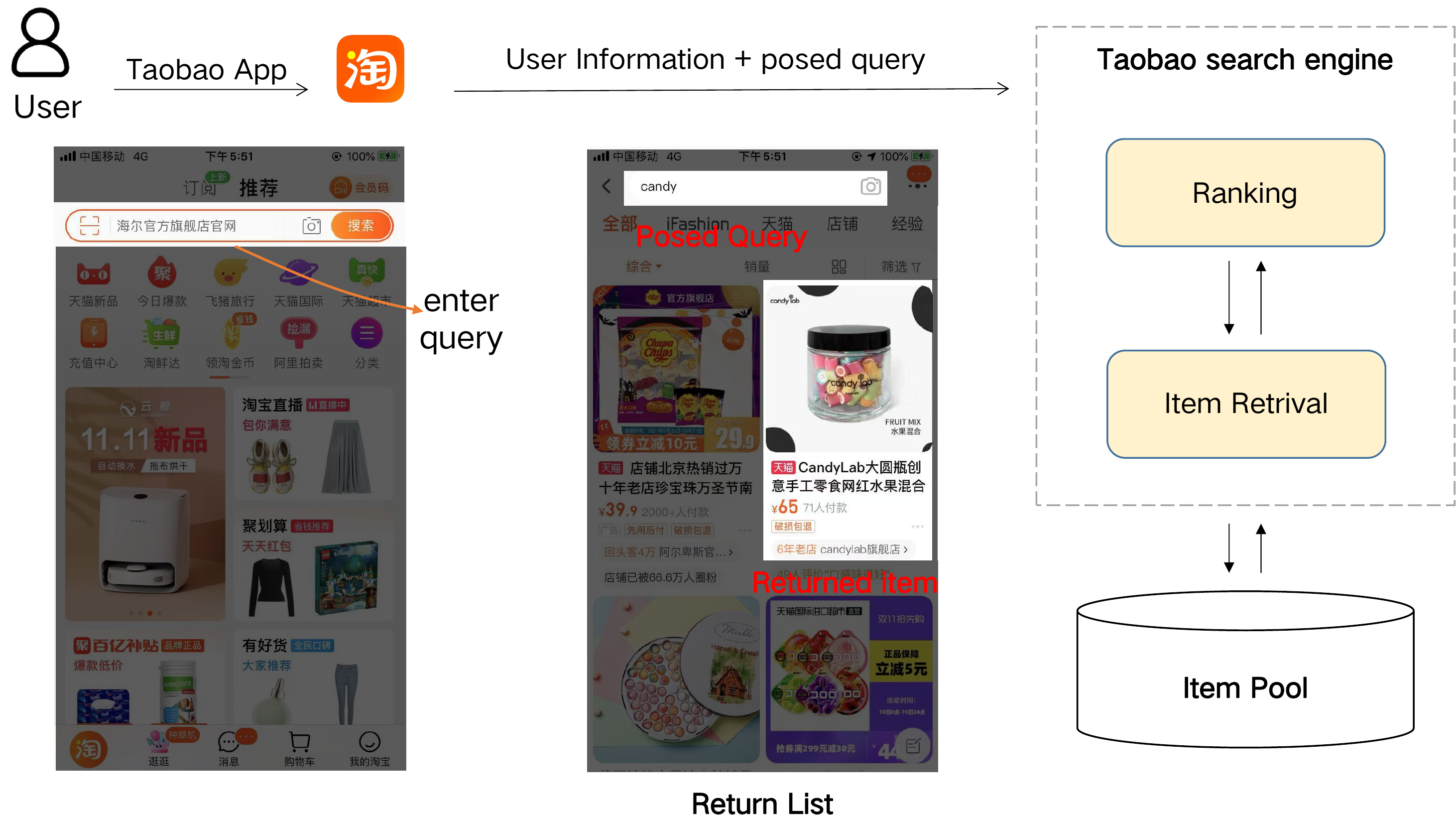}
	\caption{Workflow of Taobao’s search platform.}
	\label{fig:workflow}
\end{figure}


Our problem is set up in the context of Taobao realistic industry
setting. As illustrated in Fig.~\ref{fig:workflow}, a user can pose queries on the homepage of Taobao App. After retrieving both user information and their posed queries, the Taobao search engine analyses the provided data, and provides a return list back, which contains a small set of items from a large item pool containing hundreds of millions of item candidates. The result list is optimized towards certain key performance indicators such as platform revenue and ad sales amount, and they should satisfy the search relevance constraints. 

To balance between efficiency and effectiveness, the Taobao search engine generally applies a multi-stage search architecture for item selection. Firstly, the search engine retrieves a set of relevant items from a large item pool (Item Retrieval procedure), then ranks the retrieved items by elaborate models (Ranking procedure), and finally determines the content of return list by combining both users' provided information and ads owner's budget. 

In this paper,
we mainly focus on the retrieval stage. The goal of the proposed \sys is to work as a high-performance, scalable ad retrieval framework in the Taobao search engine. We construct relevance graphs
from users’ historical behaviors and feature similarity and aim at using the
graph model to infer the relevant and high-quality ads.

\para{Retrieval Graph.}More formally,
we encode the user-query-item relevance as a heterogeneous graph $\mathbf{G}$=\{$\mathbf{U},\mathbf{Q},\mathbf{I},\mathbf{E}$\}. 
The node types include user $\mathbf{U}$, query $\mathbf{Q}$, and item $\mathbf{I}$. We illustrate the features of each type of node we use in Table~\ref{nodeFeatures}. 


We construct two categories of edges: (1) Interaction edges, i.e., click edges and session edges (for clicks in the same session). Given a click sequence of $s$ = ($i_1, \cdots, i_m$) under a user $u$'s searched query $q$, we build interaction edges between $u$ and the searched query $q$, two adjacently clicked item $c_i$ and $c_{i+1}$, and between each clicked node $c_i$ and the query $q$. (2) Similarity-based edges, we add edges to similar nodes according to their content similarity. Specifically, we employ minHash to calculate Jaccard similarities between queries and items and use the Jaccard similarities as weights to establish similarity-based edges. This kind of edge is more informative for cold-start problems in RS, for new nodes often have sparse relations. Each edge $e=(N_1,N_2)$ $\in$ $\mathbf{E}$ denotes a connection between node $N_1$ and node $N_2$ of a certain type from the above regulations. According to the edge type of $e$, $N_1$ and $N_2$ of $e$ can be either of the same type or of different types. 



\para{Graph-based Retrieval Problem.}
Given the retrieval graph $\mathbf{G}=\{\mathbf{U},\mathbf{Q},\mathbf{I},\mathbf{E}\}$, let $f$ be a function that returns a matching score of an item $a \in \mathbf{I}$ given a query $q \in \mathbf{Q}$ of a user $u\in \mathbf{U}$, the ad retrieval problem is to determine the relevance score between an item $a'$ and corresponding user u and query q:  
\begin{equation}
a' = \arg \max_a f(u, q, a|G),  
\end{equation}
The key to this problem is to learn or extrapolate the utility function $f$ from given graph data.

\begin{table}[t]
\caption{Node Types and Features.}
\centering
{
\noindent
\renewcommand{\multirowsetup}{\centering}
\resizebox{0.95\linewidth}{!}{
\begin{tabular}{c|c} 
\toprule 
Node&Features\\
\midrule  
User&ID, Gender, Membership Level\\
\midrule  
Query&Category, Title Terms\\
\midrule 
Item&ID, Category, Title Terms, Brand, Shop\\
\bottomrule 
\end{tabular}}}
\label{nodeFeatures}
\end{table}

\section{Preliminary}
In this section, we introduce some background knowledge related to \sys. 
\subsection{Graph Neural Networks in Recommender Systems}
Since recommendation can be naturally regarded as a graph problem, Graph Neural Networks (GNNs)~\cite{zhang2021rod,miao2021degnn} has become one of the most powerful methods by applying deep learning in graphs for recommendations. Here, we briefly summarize four typical GNN frameworks which are widely adopted in the field of recommendation. 

\begin{itemize}
    \item {\textbf{GCN}}~\cite{kipf2016semi} updates the (l+1)-th convolution layer $\mathbf{H}^{l+1}$ of GCN is computed as
    \begin{equation}
    \mathbf{H}^{(l+1)}=\sigma\left(\tilde{\mathbf{D}}^{-\frac{1}{2}} \tilde{\mathrm{A}} \tilde{\mathbf{D}}^{-\frac{1}{2}} \mathbf{H}^{(l)} \mathbf{W}^{(l)}\right)\label{eqn:gcn}
    \end{equation}
    where $\tilde{\mathrm{A}}$ is the adjacency matrix with self-connections, $\tilde{\mathbf{D}}$ is the degree matrix of $\tilde{\mathrm{A}}$, and $\sigma(\dot)$ is a nonlinear activation function such as Sigmoid. $\mathbf{H}^{(l+1)}$ is the output of l-th layer of graph convolution network  ($\mathbf{H}^{0}$ is the original node feature matrix), and $\mathbf{W}^{(l)}$ is a learnable matrix for propagating features at the l-th layer. 
    User-interaction-based methods~\cite{berg2017graph,zhang2019star} follow the GCN pattern and exert mean weights on historical product node neighbors of each user. 


    \item {\textbf{GAT}}~\cite{velivckovic2017graph} and its variants\cite{velivckovic2017graph,wang2019kgat,zhang2021hyperbolic,yang2020program,xu2020probabilistic} assign weights to edges of neighbors according to the following metric:
    \begin{equation}\label{attentionWeight}\tiny
    \alpha_{v j}=\frac{\exp \left(\text { LeakyReLU }\left(\mathbf{a}^{T}\left[\mathbf{W}^{(l)} \mathbf{h}_{v}^{(l)} \oplus \mathbf{W}^{(l)} \mathbf{h}_{j}^{(l)}\right]\right)\right)}{\sum_{k \in \mathcal{N}_{v}} \exp \left(\operatorname{LeakyReLU}\left(\mathbf{a}^{T}\left[\mathbf{W}^{(l)} \mathbf{h}_{v}^{(l)} \oplus \mathbf{W}^{(l)} \mathbf{h}_{k}^{(l)}\right]\right)\right)}
    \end{equation}
    where $\mathbf{a}$ is a learnable parameter. As shown in~\cref{attentionWeight}, weights are pair-wisely calculated, despite different weights assigned for different neighbors, the weight of each edge in graphs is still fixed across different queries and users. 

\begin{figure*}[htbp]
\centering
\includegraphics[width=17cm]{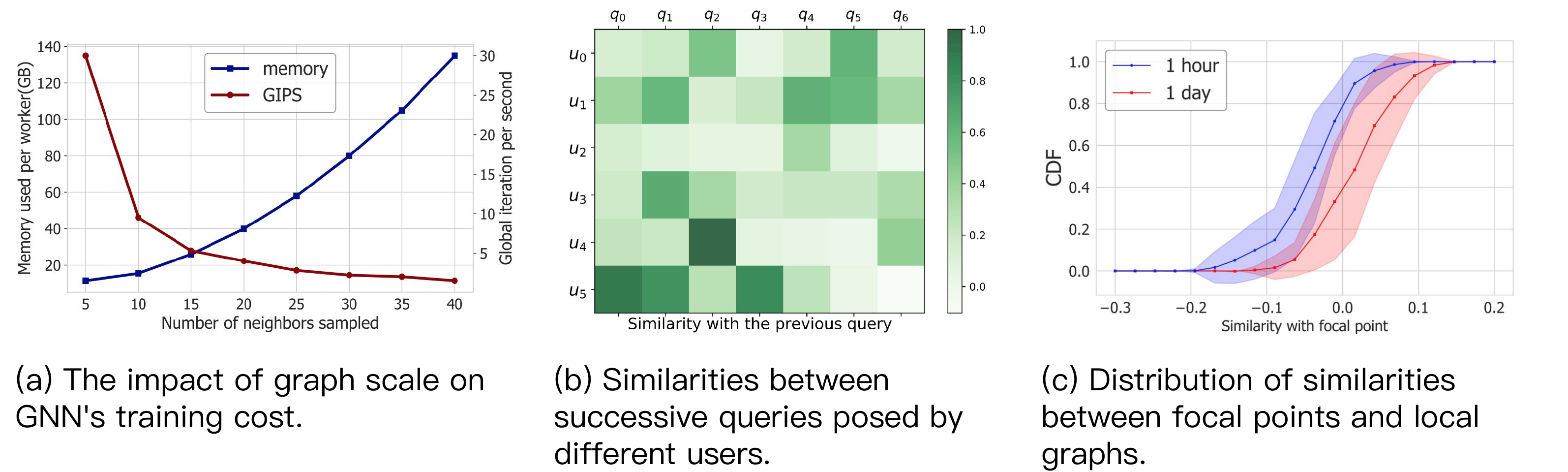}
\caption{Motivating observations of training Web-scale graphs at Taobao.}
\label{Fig.observations}
\end{figure*}

Besides, as GATs have much overhead of importance calculation, the training scale of graphs is often up to thousands of nodes. 

\item{\textbf{GraphSage}}~\cite{hamilton2017inductive} and its variants~\cite{yang2020multisage,lo2021graphsage} reduce this overhead and speed up the training by aggregating features from a set of \textit{sampled} neighbors, to make GNNs more capable of handling  graphs in RS. The major process of each convolution layer can be represented as
\begin{equation}
\mathbf{H}_{\mathcal{N}(v)}^{(l+1)}=\text { AGGREGATE }\left(\left\{\mathbf{H}_{u}^{(l)}, \forall u \in \mathcal{N}(v)\right\}\right)
\end{equation}
where $\mathcal{N}(v)$ is the sampled neighborhood of node v, and AGGREGATE is the neighborhood aggregate function. Many RS leverage this line of sampling strategies to tackle recommendation problems on large and complex real-life graphs. Note that, similar to GCN, in GraphSage each neighbor still has a fixed weight.
\end{itemize}

\subsection{CTR Prediction}
CTR prediction~\cite{mcmahan2013ad} is a core task for RS. It focuses on predicting the probability that whether a product would be clicked by users under a submitted query. With the success of deep learning, a large number of deep learning methods have been proposed for this task in the manner of learning semantic relevance between users and items. 

Particularly, twin-tower architecture has been widely used in existing works~\cite{yang2020mixed}. Each tower is trained separately to learn a certain aspect in RS (e.g., item features), and they are connected in higher layers by performing a simple operation like a dot on the outputs of both towers. As this line of methods is trained separately thus ensuring the speed of model training, twin-tower architecture is widely used in the recall phase in recommendations.

\subsection{Graph Sampling Strategies in Recommender Systems.} 
Due to the high computational complexity of directly performing GNNs on graphs in RS, GNNs in RS often apply sampling strategies to reduce high computational complexity. A general graph sampling strategy is layer sampling techniques~\cite{hamilton2017inductive,chen2018fastgcn,ying2018graph,gao2018large,lee2012beyond,zhao2020preserving}. Generally, the layer sampling methods follow the following three major steps: 1) Sample nodes or edges in a layer-wise manner as minibatch training data, 2) propagate forward and backward among the sampled nodes. The step of 1) and 2) alternate for model training. Pinsage~\cite{ying2018graph} performs uniform node sampling on the previous layer neighbors. Pinnersage ~\cite{pal2020pinnersage} performs an importance-based sampling strategy and  Multisage\cite{yang2020multisage} samples neighbors out of products' property. \sys follows this line method for downscaling web-scale graphs. 

Another important line of sampling methods is a random-walk-based sampler. Instead of layer-wise sampling, random-walk-based methods perform random walks on graphs~\cite{li2015random,perozzi2014deepwalk,eksombatchai2018pixie}, and place nodes that appear in the same walk closely together in the embedding space. Deepwalk~\cite{perozzi2014deepwalk} samples random walks from the graph and treats them as equal equivalents. Pixie~\cite{eksombatchai2018pixie} randoms edge selection to be biased based on user features.

\section{Motivation}
\para{Information Overload.}
At Taobao, the graphs for recommendation have grown rapidly to the scale of millions to billions of nodes and billions of edges. This increasingly large scale induces exponentially higher training costs of GNNs. We experimentally illustrate this by training a two-layer GCN on a real graph from Taobao with 2 million nodes and 4 billion edges and alter the number of sampled neighbors for each node to simulate the effect of growing graph sizes. The left part of Fig. \ref{Fig.observations} shows that when we sample more neighbors, the memory usage increases exponentially and the training speed (number of iterations per second) drops quickly. The training efficiency over very large graphs has become a major concern limiting the scalability of GNNs and inspired prior work on graph sampling to reduce the graph sizes.

This paper explores a new and counter-intuitive observation on such large graphs: the information in the graphs, albeit rich, is \textit{not all relevant to and helpful for each specific recommendation request}. We refer to this phenomenon as \textit{information overload}. Information overload is actually an inevitable result at such a big e-commerce platform like Taobao with the extremely massive amount of user, product, and user behavior data. We conduct measurements over Taobao data to demonstrate information overload, and find that it is manifested in the following two dimensions:

 \begin{figure*}[htbp]
\centering
\includegraphics[width=15cm]{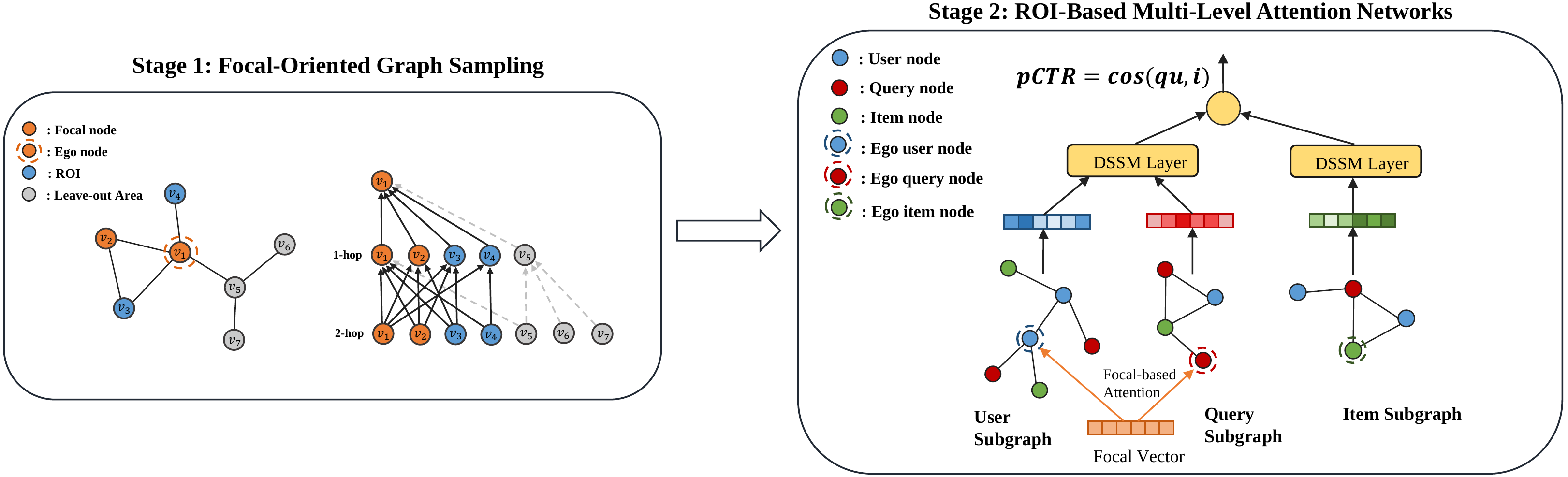}
\caption{Pipeline overview of \sys.}
\label{Fig.overview}
\end{figure*}
 

First, \textit{dynamic focal interests}. We find that the interests explicitly expressed by the users' searched queries, or \textit{focal interests}, may change quickly.
In the middle part of Fig. \ref{Fig.observations}, we show the distances between successive user queries in a few randomly selected sessions, where each $u$-$q$ pair shows the similarity between a query $q$ and the previous posed one from the same user $u$. We observe that the successive user queries within a session usually have low similarities to each other, indicating that the focal interests of users dynamically change, even in a short period. 

Second, \textit{a small relevant area in the graph to a specific focal interest}.
Given a focal interest, the historical behaviors of the user (e.g., previously clicked items) may also have low relevance to this latest interest. To evaluate this effect in both users' short-term interests and long-term interests, we build two graphs from user behavior logs in Taobao of 1 hour and 1 day, respectively. For each graph, we randomly select 10 user nodes; for each user, we randomly select a query node, and take both the user and the sampled query node as focal points, then calculate the cosine distances between the focal points and all the clicked items by the user. The 
right part of Fig. \ref{Fig.observations} shows the CDF of the distances between the focal points and the interaction-based users' local graphs, where the curves and the colored areas show the mean values and standard deviation of the results across the selected users, respectively. The result shows that the similarities are generally low---most of them fall in the range of $[-0.1, 0.1]$, and roughly 80\%/40\% are lower than 0.0 in the 1-hour/1-day graph, which indicates that the value of similarity between  specific focal points and a certain local graph is small. 


\para{Regions of Interests.} The characterization of information overload above has deep implications for GNN-based recommendation: for each user-query pair as a focal pair, there is a distinct area in the graph that is smaller but of higher relevance to the query and hence acquiring  higher importance for this recommendation request. We refer to such an area as the \textit{Region of Interest} (ROI) of this focal interest. Computation over the other area outside the ROI not only has little necessity but also may distract the GNN model and impact the recommendation quality negatively.

The key idea of \sys is to dynamically identify and focus on the ROI of each focal interest. The benefits of focusing on ROIs are multi-fold. First, it can eliminate unnecessary information interference and fully leverage users' long-term interests to improve the quality of recommendations. Besides, it reduces the computational cost by downsizing the graph, while still preserving the more important part of the graph.
Moreover, it can guide the GNN model to learn more from the more relevant information and mitigate the impact of distracting information to improve the recommendation quality. 
This motivates us to design \sys, a system exploiting ROIs to deliver both higher efficiency and higher quality of the GNN.

\section{\sys Design}

\subsection{Pipeline Overview}\label{sec:overview}
\sys is a system for high-efficiency and high-quality recommendations on web-scale graphs.
The overall architecture and workflow of \sys are illustrated in Figure \ref{Fig.overview}.
Built upon the key concept of ROI, \sys operates in two major stages as follows:

\para{Graph sampling for ROI construction.}
For each recommendation request, \sys first identifies the nodes representing the user's intention as the focal points in the graph. Based on the similarities with the focal points, \sys further samples a neighborhood region with high relevance to the focal to construct the ROI sub-graph from the huge graph that contains users' long-term interests and will be fed into the multi-level attention module.

\para{Multi-level attention over the ROI.} We use three separate models for learning embeddings of users, queries, and items, respectively. Similar to the mainstream twin-tower model architectures \cite{yang2020mixed}, one tower handles a combination of user and query embeddings, and the other one handles the item embeddings. 
Given the ROI, \sys first constructs a focal vector of selected focal points. We first retrieve the embeddings of the focal points from embedding tables separately, then we perform space mapping on focal points of different types into the same latent space. After this, We directly sum up the processed focal points' representations to a focal vector. With the focal vector, in each of the user and query models, \sys introduces a multi-level attention mechanism to guide the GNN to reweigh different parts of the graph according to their relevance to the focal vector, further focusing on the more relevant information in the ROI. This attention mechanism is performed at multiple levels, from features in the nodes, to edges, and to the semantic difference of different types of nodes, to capture the multiple aspects of a heterogeneous graph. Finally, the trained embeddings are used for downstream tasks like CTR prediction.

\subsection{Focal-Selection}\label{sec:pruning}

To construct an ROI, we first need to identify the \textit{focal points} in the graph that are most representative of the user's intention.
Recall that in search recommendation, users first pose queries out of timely intention, and then click on items from the retrieved list by the RS. User behaviors under this context can be summarized as a tuple \{$u_k,q_k,i_k$\}, which represents that user $u_k$ searched query $q_k$ and clicked item $i_k$ under $q_k$. We refer to each node in this tuple as the ego node of this recommendation request in the user, query, and item models, respectively.

\textit{\sys assigns the pair \{$u_k,q_k$\} to each node as focal points}. We select these two nodes because $u_k$ contains important personalized user information and $q_k$ contains the strong and explicit intention of the user. We do not include $i_k$ in the focal points because users may have scattered interests in products under the same category, e.g., in different designs or styles of products, and may lead the overall results towards an unexpected bias. Therefore we do not restrict the focal point to this only item to avoid deviation introduced by certain items.

While during online serving where we need to provide low latency for thousands of queries per second, applying user and query information to the item side (i.e., assigning \{$u_k,q_k$\} as the focal of $i_k$) would be time-consuming. To meet the time cost requirement, we only deploy \sys on the user-query side and use a base item model to extract item embeddings. To align offline training and online deployment, we use the same settings in offline training as online training. Specifically, we use the aforementioned user-query pairs as focal points to train the embeddings for ego nodes on the user-query side. 

\subsection{Focal-biased Graph Sampler}
As focal points represent users' timely intentions, \sys further leverages the assigned focal points to sample a neighborhood region with significant relevance to the focal as the ROI. This sampling further reduces the amount of computation and also filters out noisy examples which may deteriorate the recommendation quality but are prevalent in real-world implicit feedback. For example, a user's experience on purchasing household items may have less relation with how she chooses luxuries. Besides, outdated historical user behaviors may also become noises as user interests change over time. 




To this end, instead of performing random sampling, we bias the sampling of \sys in a focal-oriented way. We give an example of the proposed Focal-biased Graph Sampler in Fig.~\ref{Fig.overview} (Stage 1). For the ego node $v_1$, we first assign both $v_1$ and $v_2$ as its focal points.  To preserve localized neighbors most relevant to focal points of the ego node, we calculate the relevance to the focal point for the neighbors of the ego node $v_1$.
We elect $v_3$ and $v_4$ as an ROI for $v_1$ under these focal points for higher relevance scores and abandon $v_5$ and its neighbors for they are less relative with current focal. The resulting graph sampling path (2-hop) is also shown in the figure.

We design a formula to calculate the focal-relevance score $e_{ij}$ for ego node $V_i$ and its neighbor node $V_j$ under focal points $\mathbf{c}$ as:
\begin{equation}
    e_{i j}=\frac{\mathbf{F_{c}} \cdot \mathbf{F_{j}}}{\|\mathbf{F_{c}}\|^{2}+\|\mathbf{F_{j}}\|^{2}-\mathbf{F_{c}} \cdot \mathbf{F_{j}}}
\end{equation}
where the $\mathbf{F}$s are node features. We directly sum up embeddings of focal points in $\mathbf{c}$ as $\mathbf{F_{c}}$. $e_{i j}$ will be larger if $V_j$ is more relevant to the focal points $\mathbf{c}$. 
Note that, since $\mathbf{F_{c}} \cdot \mathbf{F_{j}}$ is positively relative to the similarity degree between $\mathbf{F_{c}}$ and $\mathbf{F_{j}}$, certain nodes with higher similarity scores can be distinguished to form the Region of the Interests, and filter non-relative information out from users' historical behaviors logs. This process offers relevance guarantee between retrieved products and focals, as the selected parts are either highly relevant to the posed query, or accord with specified user tastes. Following the designed principle that nodes similar with focals should be assigned with higher scores, eq.(5) can be replaced with other relevance score equations like cosine distance. 

We score all neighbors of node $V_i$ according to the above equation, and sample neighbors in a top-k manner (k is a hyper-parameter) to produce the ROI subgraph for the ego nodes.

\begin{figure}
	\centering
	\includegraphics[width=0.95\columnwidth]{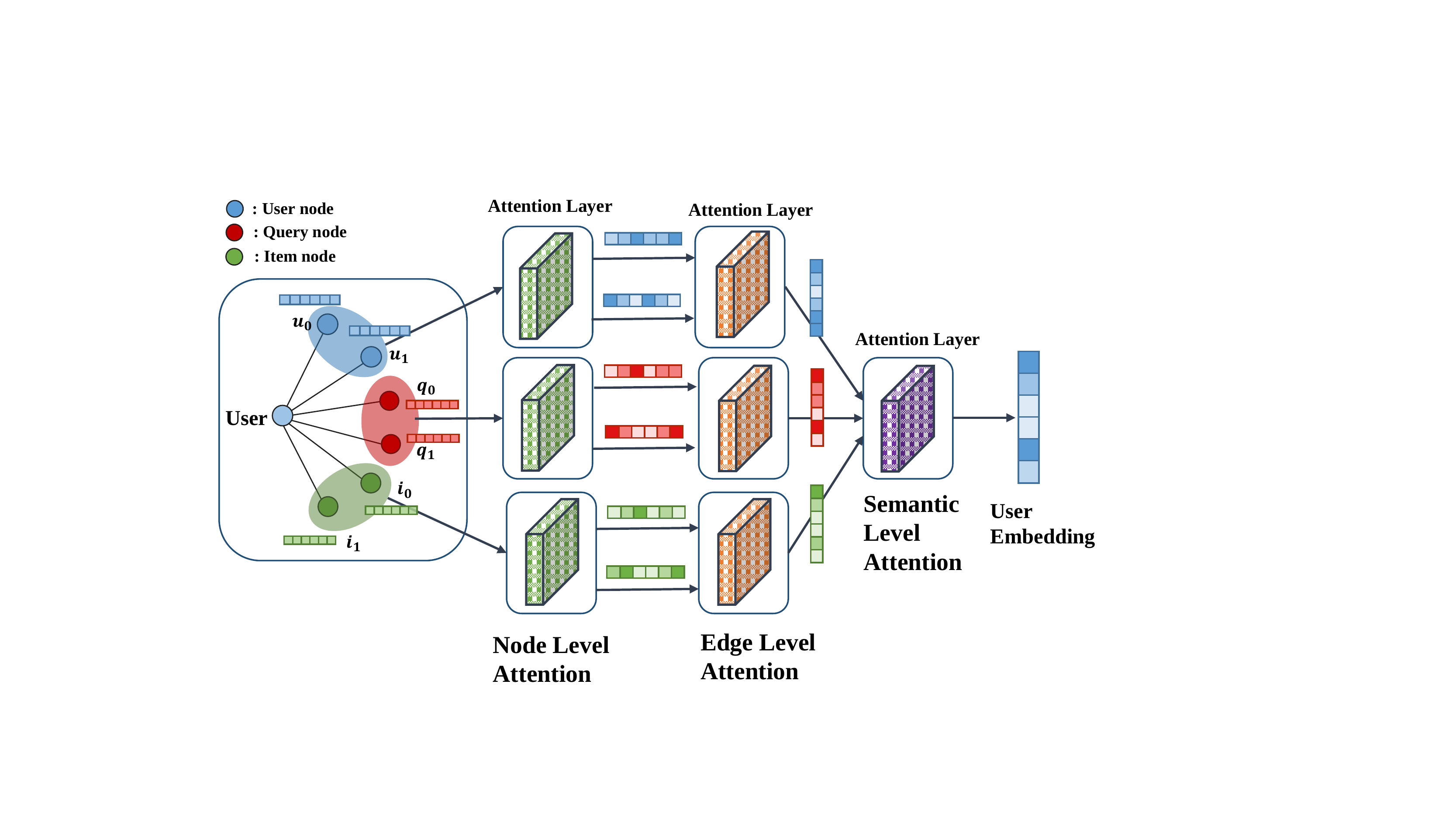}
	\caption{Overview of the Multi-Level Attention Module.}
	\label{fig:attentionFramework}
\end{figure}

\subsection{ROI-based Multi-Level Attention }\label{sec:attention}

The construction of ROIs is the first, and a coarse-grained step to filter out irrelevant information from user historical graphs since it can only sample nodes while preserving all of their features and relations. However, the fine-grained information within the sampled ROI nodes is still rich and sophisticated, showing a great difference in the relevance to the focal interest. To further address the information overload problem within the coarse-grained ROIs, \sys introduces an attention module to guide the GNN model to focus on information that is more relevant to the focal interest.



The major design consideration of this attention module is the interplay between information overload and data heterogeneity, i.e., to consider the information overload problem from different aspects of a heterogeneous graph:
(1) the \textit{nodes} (multiple types of nodes with abundant feature information); (2) the \textit{edges}, constructed from different types of behaviors (e.g., click, co-concurrence) or similarities; (3) the above two combined, i.e., the \textit{semantics of edges between different types of nodes}---for example, query-item relations are usually more important than item-item relations because items recommended under a specific query should have the same category with this query.

\sys introduces a \textit{multi-level attention} module to extract focal-relevant information from the ROI, where the multiple levels mitigate information overload in all the above aspects in a heterogeneous graph. To the best of our knowledge, \sys is the first GNN system that simultaneously concerns all three levels of graph heterogeneity. 
Fig. \ref{fig:attentionFramework} shows the overall architecture of this module. We take user nodes as an example, and nodes of other types follow the same attention process. This module takes the heterogeneous graphs and the corresponding focal vectors as inputs and outputs node embeddings for downstream tasks. According to the relevance to the focal vector, this attention module calculates (1) weights of features of each node, then (2) weights of edges for aggregating features, and finally (3) the importance of different types of neighbor nodes and combines them to obtain unified node embeddings. 

\para{Feature Projection.}
Different features of a node may have different relevance to a specific query. For example, when a user poses a general query like ``dress'', the feature of the user's age may take much effect, while some other information like place of residence may have less importance.

\sys identifies more important feature information by performing \textit{feature-level attention} with respect to a particular focal interest.
Specifically, after space mapping from the original features, we assume that each node in the graph is represented by a series of latent vectors represented as $H ={ H_1,H_2,\dots,H_n }$, where n is the number of features for nodes. Each $H_i$ in $H$ represents embedding of the corresponding feature. 
We calculate weights for the features' latent representations with respect to given focal points $\mathbf{C}$:
\begin{equation}
W_c\left(H, \mathbf{C} \right)=\operatorname{softmax}\left(\frac{H \mathbf{{C}}}{\sqrt{ d}}\right) H
\end{equation}
where $W_c \in \mathbb{R}^{1 \times n}$ denotes the focal-oriented attentive weights for feature latent vectors. With the assigned weights, we then learn transformed features as follows:
\begin{equation}
Z_c\left(H, W \right)=H \odot W_c
\end{equation}
where $\odot$ is used to multiply the weights on feature latent vectors. 
This way, we amplify the influence of focal-relevant node features and decrease the importance of less relevant ones. 

\para{Edge Reweighing.}
After the feature-level attention, we need to further aggregate the transformed features from neighbors of the same type to the ego nodes, to exert neighbor influence from a certain party on the ego node. We believe that nodes are connected with different intentions, thus have different edge relations' importance, which cannot be simultaneously captured by a single embedding, and the importance of neighbors of the same type may  vary across different focal points considered in the aggregation. The constrain of reweighing within the same type make neighbors of an ego node fairly comparable. 




To select important edges and distinguish the influence of different types, \sys introduces \textit{edge-level attention} to identifying edges more relevant to the focal nodes of the same type when aggregating features through the edges. 
Similar to the feature-level attention, we calculate the weights of edges as follows:

\begin{equation}
    e_{ij}=\frac{\exp \left(\sigma\left(\mathbf{a}^{T} \cdot\left[(Z_{f i}\|Z_{f j})\| \mathbf{Z_{f_c}}\right]\right)\right)}{\sum_{k \in \mathcal{N}_{ti}, t \in \mathcal{T}} \exp \left(\sigma\left(\mathbf{a}^{T} \cdot\left[(\mathbf{Z_f}_{i} \| \mathbf{Z_f}_{k})\| \mathbf{Z_{f_c}}\right]\right)\right)}
\end{equation}
where $e_{i j}$ is the focal-oriented edge weights under focal points $\mathbf{c}$, $\mathbf{Z_{f_c}}$ is the output of feature projection of $\mathbf{c}$, $\mathcal{N}_{ti}$ is $i$'s neighbors of type $t$, $\mathcal{T}$ is all types in graphs, and $\mathbf{a}$ is a learnable weight vector. 

When performing edge-level propagation, we calculate focal-oriented embeddings for node $j$ at edge level in the following format:
\begin{equation}
    E_{jT}=\sum_{k \in \mathcal{N}_{T_j}}  e_{jk}*Z_{k}
\end{equation}
where $\mathcal{N}_{T_j}$ is the localized neighbors of type $T$ of node $j$, and $Z_k$ is the feature embedding of node $k$. 

\para{Semantic Combination.}
When aggregating features from neighbors, there is another dimension of heterogeneity to consider: edges between ego nodes and different types of neighbor nodes with different semantics, and describe user behaviors in different aspects. 
    We propose a semantic combination to handle multipartite networks at the semantic level. 

We aim to learn the weights of types of neighbor nodes according to ego node $i$'s type $t$ = ($t_1,t_2 \dots t_n$), and each weight of type $k$'s importance to ego node i is composed as follows:
\begin{equation}
    t_{k}= cosine\_sim(C_i, E_{ik})
\end{equation}
where $cosine\_sim(,)$ is the cosine distance function. $C_i$ is the feature-level embedding of node $i$, and $E_k$ is the edge embedding of $i$'s localized neighbors of type $k$. 
We aggregate the edge embeddings of different types by using $t$ as weights:
\begin{equation}
    H_{i} = \sum_{k \in \mathcal{T}}  E_{ik}*t_{k}
\end{equation}
where $k$ is the neighbor type.



Once obtaining the final embeddings from $H_{i}$, we pass these embeddings through a simple MLP-based twin-tower model to predict ratings from users to items. 




\section{\sys Implementation}

\begin{figure}
	\centering
	\includegraphics[width=0.95\columnwidth]{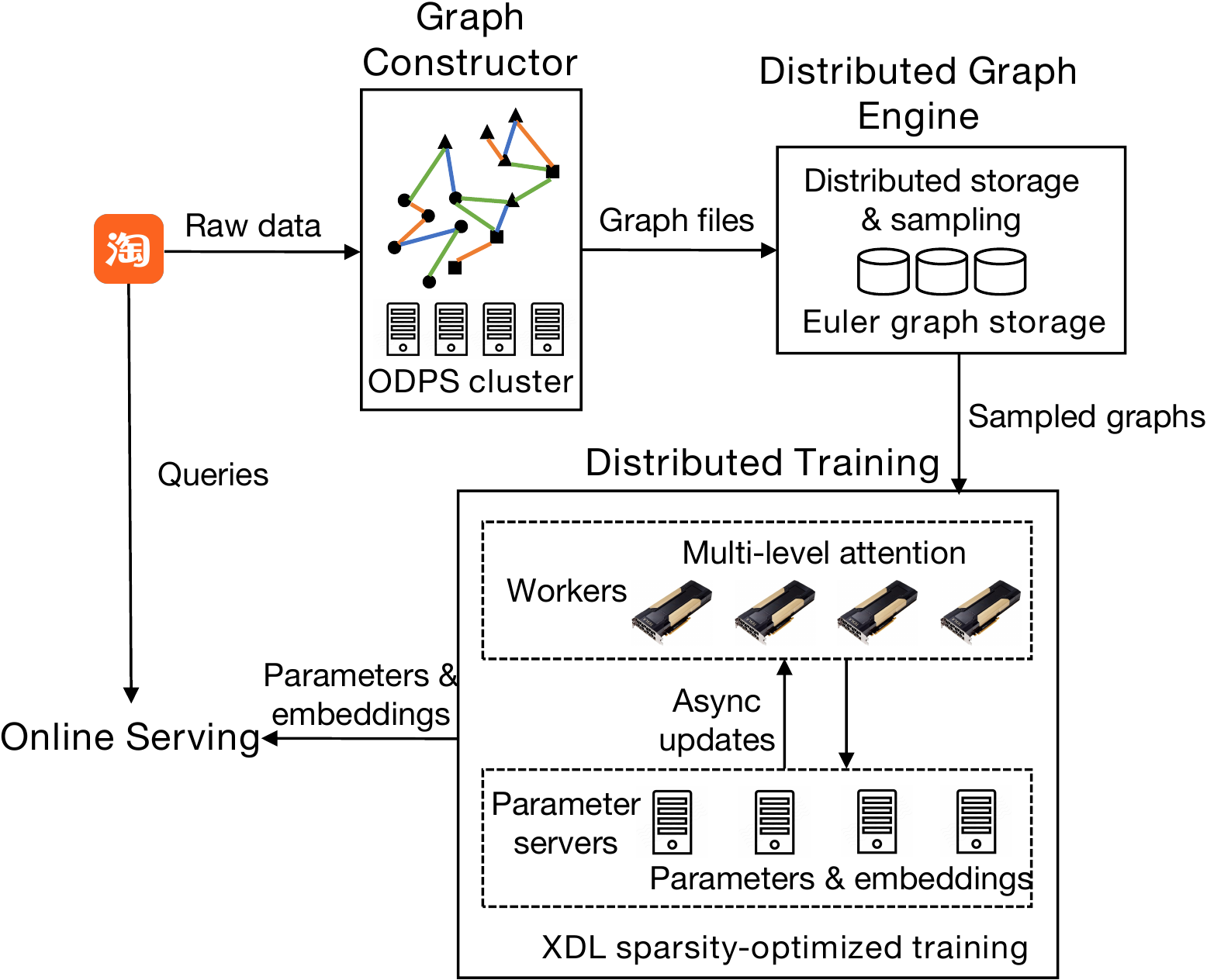}
	\caption{Architecture of the system implementation.}
	\label{systemImpl}
	\vspace{2mm}
\end{figure}

\sys is implemented as a scalable distributed system to facilitate training over web-scale data. It builds upon an open-source toolchain of distributed computing frameworks developed at Taobao, as illustrated in Fig. \ref{systemImpl}.

The implementation of \sys comprises three major components: (i) a graph generator that transforms the raw data (i.e., the user behavior logs) to graphs, (ii) a distributed graph engine for storing, processing, and sampling the graphs, and (iii) a distributed deep learning training framework over sparse data for efficient training of the embeddings. All of these combined, \sys can handle graphs with billions of nodes efficiently.

\para{Graph generator.}
The graph generator is implemented in the open-source distributed data analytics framework ODPS (Open Data Processing Service)~\cite{ODPS}. The customer-platform interactions are recorded in real-time and logged into ODPS. Following the user-designed rule, ODPS parses the logs and constructs heterogeneous graphs from the data. The graphs are stored using compact binary-format files. These files are saved into HDFS to be accessed by the graph engine.

\para{Distributed graph engine.}
Our graph engine is implemented using the open-source graph deep learning framework Euler~\cite{Euler} for the distributed storage and sampling of the graphs. The graph engine loads the graphs from HDFS (Hadoop Distributed File System) and stores them in memory.
To scale to very large graphs efficiently, the graph engine stores the graphs in a distributed manner: a graph is partitioned into multiple shards for higher storage capacity, and each shard is replicated onto multiple servers for higher aggregate throughput. On each server, we also implement efficient metadata and data structures to improve the storage and sampling performance. We use an Alias Table \cite{alias} to implement the adjacency list to achieve constant-time graph sampling independent of the graph size. We use compact data structures to store different types of features in heterogeneous graphs with high memory utilization.

\para{Distributed training.}
\sys performs large-scale distributed training using the open-source deep learning framework XDL \cite{XDL}.
\sys trains the model using a worker-PS (Parameter Server) architecture. \sys partitions and stores the model parameters and the embeddings on multiple parameter servers. In each training iteration, each worker reads a mini-batch of data sampled from the graph via the graph engine and retrieves the parameters and the embedding tables from the parameter servers. The workers will push the updates back to the parameter servers after each iteration. Considering the model parameters are highly sparse, i.e., there is a low probability of conflict of concurrent parameter accesses across different workers, the workers retrieve and update parameters \textit{asynchronously} to improve training efficiency on large models.

Targeting sparse data, \sys incorporates a series of optimizations to improve training performance. On the worker side, observing that training our model involves IO-intensive operations for both the input data and the embeddings, \sys overlaps the threes stages of reading subgraphs, reading embeddings, and the training computation in a fully asynchronous pipeline to avoid IO bottleneck. \sys also employs various computation kernels specially optimized for sparse data. On the PS side, \sys develops techniques to improve the efficiency of exchanging sparse data, such as network communication optimizations for small packets and dynamic load balancing strategies for better partitioning of the parameters and the embeddings across PSes.

After training, the representations are fed to an efficient Approximate-Nearest-Neighbors search module (ANN) to generate the inverted index for online serving. When a customer poses a query to Taobao app, the online product retrieval module retrieves items by searching over the inverted index in iGraph system\cite{iGraph}. 

\section{Experiments}
To validate the effectiveness of \sys, we conduct a comprehensive suite of experiments on both open benchmark datasets and real-world networks from Taobao historical logs, including offline experiments and production A/B tests. Our code is publicly available. 

\subsection{Experimental Setup}
\label{dataset_description}
\para{Dataset Statistics.} We evaluate \sys on one open benchmark MovieLens and three industry datasets from Taobao. 

For the open benchmark MovieLens, we dispose of the original data for constructing heterogeneous graphs and focal tuple pairs out of our needs. The input of our model is a triple pair of {user, tag, movie}, the binary output indicates whether the user has interacted with the movie under the given tag. We split 80\% of the dataset as the training set and 20\% as the test set. We construct the graph based on the MovieLens 25M dataset, and our heterogeneous graph contains three types of nodes: 209171 movie nodes, 162541 user nodes, and 1128 tag nodes. The user-movie edges are constructed based on the users' rating for movies, and the movie-tag edges are based on the relevance scores which are preprocessed using machine learning algorithms. For each movie node, we select the top 5 tag nodes in relevance scores as neighbors.

For the industry datasets, \sys collects normal users' daily interaction information when using the Taobao app, and constructs heterogeneous graphs with the user, query, and item nodes. 
Following the graph construction rules in Section~\ref{sec:graph}, we construct three different scale graphs to evaluate \sys's effectiveness, respectively: 
\label{industry_datasets}
\begin{itemize}
\item \textbf{million-scale graph} 1-hour data from Taobao, including about 2 million nodes (1 million item nodes, 0.5 million query nodes, 0.5 million user nodes) and 400 million edges (35\% are item-item edges, 30\% are query-item edges). 

\item \textbf{hundred million-scale graph} 12-hour data from Taobao, including about 140 million nodes (7 million item nodes, 40 million query nodes, 90 million user nodes)  and 30 billion edges (75\% are user-user edges).

\item \textbf{billion-scale graph} 7-day data from Taobao, including about 1.2 billion nodes (570 million item nodes, 250 million query nodes, 340 million user nodes)  and 260 billion edges (70\% are user-user edges).
\end{itemize}

For each dataset, we use 90\% of the data for training and 10\% for testing.

\para{Baselines.} We compare \sys with the following baseline methods: 
\label{baseline_methods}
\begin{itemize}
    \item GraphSAGE~\cite{hamilton2017inductive}: GraphSAGE is an inductive framework that leverages node attribute information to efficiently generate representations on previously unseen data.
    \item Global Context Enhanced Graph Neural Networks (GCE-GNN)~\cite{wang2020global}: GCE-GNN is a framework that exploits item transitions over sessions for better inferring the user preference of the current session. 
    \item Factor Graph Neural Network (FGNN)~\cite{zhang2019factor}: FGNN is a novel model which collaboratively considers the sequence order and the latent order in the session graph for a session-based recommender system.
    \item Short-Term Attention/Memory Priorit (STAMP)~\cite{liu2018stamp}: STAMP is a model which is capable of capturing both users’ general interests and current interests.
    \item Multi-Component Graph Convolutional Collaborative Filtering~\cite{wang2020multi}: MCCF is a novel Multi-Component graph convolutional Collaborative Filtering (MCCF) approach to distinguish the latent purchasing motivations underneath the observed explicit user-item interactions.
    \item Heterogeneous Graph Attention Network (HAN)~\cite{wang2019heterogeneous}: is a heterogeneous graph neural network based on hierarchical attention, including node-level and semantic-level attentions.
    \item PinSage~\cite{ying2018graph}: PinSage is a method that combines random walks and graph convolutions to generate embeddings of nodes  that incorporate both graph structure as well as node feature information.   
    \item PinnerSage~\cite{pal2020pinnersage}: PinnerSage is an end-to-end recommender system that represents each user via multi-modal embeddings and leverages this representation of users to provide personalized recommendations. 
    \item Pixie~\cite{eksombatchai2018pixie}: Pixie is a scalable graph-based real-time recommender system that was developed and deployed at Pinterest. 
\end{itemize}

Among them, GraphSage, PinSage, PinnerSage, and Pixie own self-developed graph downscaling strategies for handling web-scale data. 

\para{Evaluation Metrics \& Parameter Settings.} 
The metrics used in our experiments are:
\begin{itemize}
  \item AUC (Area Under Curve): AUC is computed in a separate dataset with human-labeled relevance for query-item pairs. A higher AUC here indicates better retrieval relevancy. 

  \item Hitrate@K: we use hit-rate to measure the model performance. The hit-rate metric we use is calculated as $\textsf{hit-rate}= \frac{1}{M} \sum_{i=0}^{M-1} \textsf{current-state}$, where M is the number of item sampled, and $\textsf{current-state}$ indicates whether the current clicked items are provided by Recommender Systems in the top-$k$ retrieved list. In our experiment, $k$ is separately set to be 100, 200, and 300. 

  \item RPM (Revenue Per Mille): the most important online revenue measure metric, which stands for revenue per mille, and measures incomes from users' s click on advertisements (sponsored items) in RS. This metric is calculated by $\textsf{RPM} = \frac{\text{Ad revenue}}{\text{\# of impressions}} \times 1000$. 
  \item CTR (Click-Through Rate): Online metric measuring click-through rate's lift, calculated by $\textsf{CTR} = \frac{\text{\# of ad clicks}}{\text{\# of impressions}}$. 
  \item PPC (Price Per Click): Online metric measuring paying for clicks in RS. 
\end{itemize}



\para{Experimental Settings.} 
\label{hyperparameter_setting}
For Taobao datasets, \sys and all graph-based baselines, we use neighbors within two hops for aggregation. For MovieLens, \sys and all graph-based baselines use one-hop neighbors for aggregation. 

We use OpenBox~\cite{li2021openbox} or follow the original papers to get the optimal hyperparameters of our baselines. For MCCF and FGNN, regulation loss weight is set to 5e-7. For Graphsage, the learning rate is set to 0.05. For all the methods, we set the batch size to 1024.

For \sys, we apply focal cross-entropy loss and set the focal weight to 2. For \sys, we set the regulation loss weight to 1e-6, the learning rate to 0.1. We set the hidden size of the graph node embedding to 128.  For AUC and Hit-rate metric test, we set the number of sampling neighbors in each layer to 10 and trained for 5 epochs with SGD, using the Adam optimizer. 

\para{Experimental Environment.} 
We conduct the experiments using the XDL framework. Our model is trained using CPU and we use TensorFlow 1.12. For distributed training, we use 1000 workers and 40 parameter servers (PS), each worker and PS has 15 CPU cores. In addition, we use 10 Euler severs,  each Euler server has 15 CPU cores. In this setting, it takes about 10 hours to process 1.5 billion input data.
Our code is publicly available on https://github.com/lovelyhan/zoomer.

\subsection{Offline Experiments}
To evaluate \sys 's effectiveness, we compare \sys with baselines on one public dataset MovieLens, and a million-scale industry graph from users' historical logs of Taobao. 

\begin{table}[t]
\caption{~\sys benchmaketing results on MovieLens.}
\centering
{
\noindent
\renewcommand{\multirowsetup}{\centering}
\resizebox{0.6\linewidth}{!}{
\begin{tabular}{c|cccc}
\toprule
\textbf{Models}& \textbf{AUC}&\textbf{MAE}&\textbf{RMSE}\\
\midrule
GCE-GNN & 91.70 & 0.3225 & 0.4339  \\
FGNN & 90.72 & 0.3140 & 0.3742  \\
STAMP & 88.07 & 0.3590 & 0.3961 \\
MCCF & 91.92 & 0.4301 & 0.4369 \\
HAN & 90.55 & 0.3449 & 0.3961 \\
\midrule
\sys& \textbf{93.79}  & \textbf{0.3014} & \textbf{0.3760} \\
\bottomrule 
\end{tabular}}}
\label{end2end-open}
\end{table}

\begin{table}[t]
\caption{AUC and Hit-rate for \sys and baselines on Taobao-industry graph.}
\centering
{
\noindent
\renewcommand{\multirowsetup}{\centering}
\resizebox{0.98\linewidth}{!}{
\begin{tabular}{c|cccc}
\toprule
\textbf{Models}& \textbf{AUC}&\textbf{Hitrate@100}&\textbf{Hitrate@200}&\textbf{Hitrate@300}\\
\midrule
GCE-GNN & 68.3 & 0.23 & 0.31 & 0.43 \\
FGNN & 64.2 & 0.22 & 0.38 & 0.52 \\
STAMP & 69.6 & 0.30 & 0.45 & 0.56\\
MCCF & 64.6& 0.22 & 0.38 & 0.52\\
HAN & 70.3 & 0.25 & 0.36 & 0.49\\
PinSage & 68.0 & 0.23 & 0.33 & 0.45\\
GraphSage & 68.2 & 0.25 & 0.36 & 0.47\\
PinnerSage & 69.1 & 0.28 & 0.38 & 0.50\\
Pixie & 69.5 & 0.27 & 0.40& 0.53\\
\midrule
\sys& \textbf{72.4}  & \textbf{0.35} & \textbf{0.48} & \textbf{0.58} \\
\bottomrule 
\end{tabular}}}
\label{end2end-industry}
\end{table}

\para{Results on MovieLens.} To illustrate \sys's effectiveness, we compare \sys with the other 5 baselines on open benchmark MovieLens. This experiment focuses on the GNN baselines without heuristic sampling strategy, because the MovieLens dataset lacks side-information these methods need (e.g., interaction frequency between nodes). The result is shown in table~\ref{end2end-open}. We could see that \sys achieves the best performance and outperforms the best baseline by nearly 2\% AUC lifting, demonstrating \sys 's capability in capturing the most important information in given data. 

\para{Results on Taobao Graphs.} We further compare \sys with all other baselines on the Taobal dataset. The results in Table~\ref{end2end-industry} show that \sys again 
achieves the best performance by a significant margin on every metric, outperforming the top baseline by 2.1\% absolute in terms of the AUC, and also 0.1 hit-rate of average hit-rate@K.

Among the baselines, HAN is the most similar to us because
it also learns the importance of data from both edge-level and semantic-level by the attention mechanism. The key difference is that HAN does not consider dynamic user interests. \sys performs better than HAN on all the metrics, demonstrating the effectiveness of the ROI-based multi-level attention mechanism in identifying knowledge relevant to user intentions and mitigating information overload.
While compared with other graph embedding baselines like GCE-GNN, \sys's improvement shows the priority of focusing on ROIs from massive user data. 


\subsection{Ablation Study}
To measure the impact of each component in the ROI-aware multi-level attention mechanism on test AUC performance, we test \sys while disabling one component at a time. We evaluate \sys as: 
\begin{itemize}
    \item \textbf{GCN}:  mean-pooling aggregation in the edge-level;
    \item \textbf{\sys-FE}: \sys with semantic combination replaced by mean pooling.
    \item \textbf{\sys-FS}: \sys with edge reweighing replaced by mean pooling. 
    \item \textbf{\sys-ES}: \sys with feature projection replaced by the original feature;
    \item \textbf{\sys} with all optimizations presented in this paper.
\end{itemize}

\begin{figure}[tpb!]
    \centering
    \includegraphics[width=0.7\linewidth]{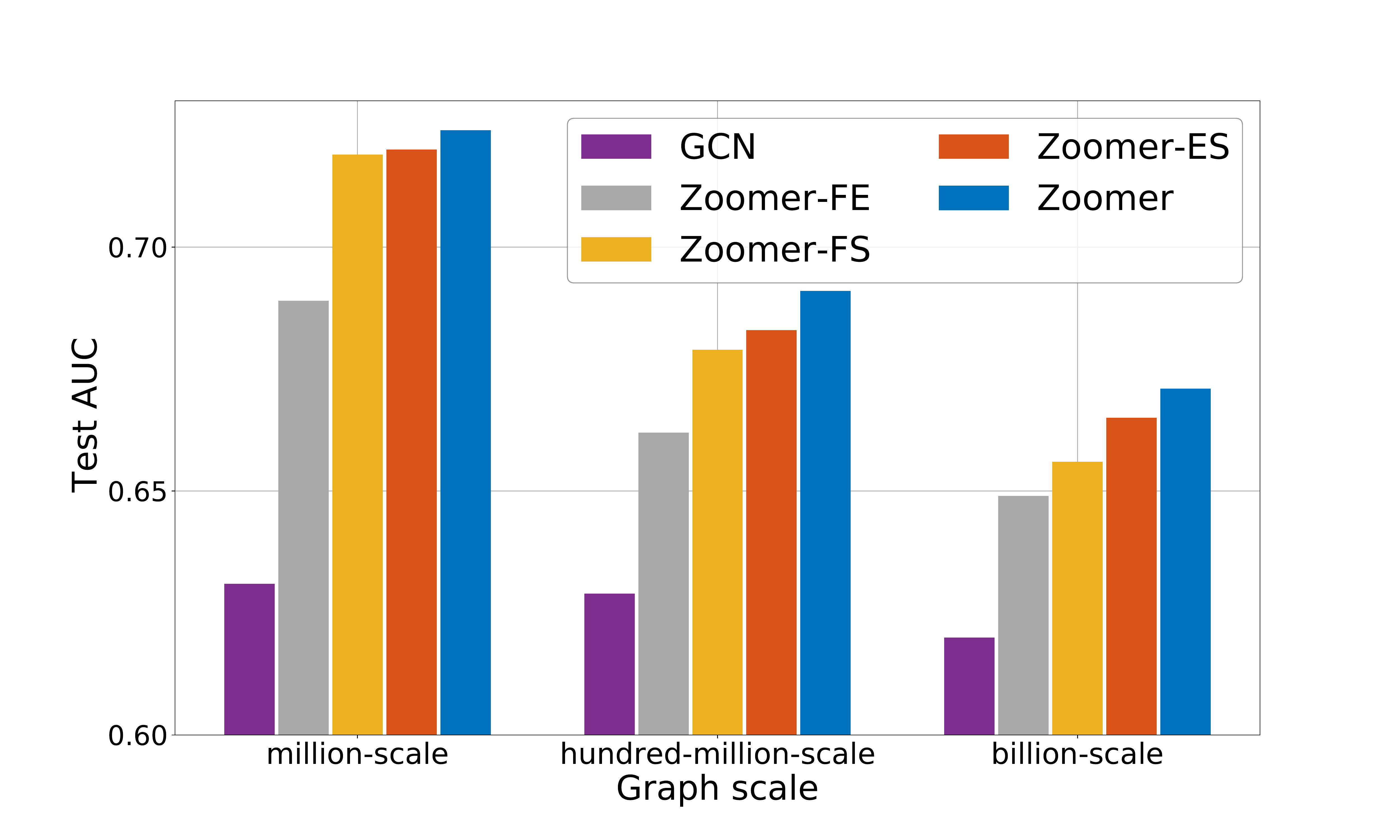}
    \vspace{-4mm}
    \caption{Ablation study results.}
    \label{ablation}
    \vspace{-4mm}
\end{figure}

This experiment uses the three Taobao graphs of different scales described in~\ref{industry_datasets} to assess the effectiveness of mitigating information overload on huge-scale data of the above techniques. We report the result in Fig.~\ref{ablation}. Compared with plain GCN, we could see that attention at every level brings improvement to the AUC, showing the effectiveness of extracting available information. 

Especially, we found that if we remove the semantic-level part from \sys, the performance drops more quickly compared with other variants. The only difference between this variant and the complete version is whether we take the importance of heterogeneity into account. Without consideration on this side, the results drop quickly, demonstrating that the heterogeneity factor is a key element of information overload in web-scale user interaction graphs. Besides, among all variants, we find that \sys-ES brings the most gain in all three Taobao graphs of different scales, illustrates that inter-relations between nodes are more important than node inner-relations. Beyond the above observations, we find that performance on larger graphs is worse than that on the small ones. One major cause for this is that larger graphs need more training steps for better performance which exceeds our expected training cost, thus leading to sub-optimal states. Besides, it implies that the additional information in the larger web-scale  graphs does not necessarily lead to higher model performance under a limited training budget. 


\begin{table}[t]
\caption{End-to-end comparison to Production Base Model \sys.}
\centering
{
\noindent
\renewcommand{\multirowsetup}{\centering}
\resizebox{0.5\linewidth}{!}{
\begin{tabular}{ccc}
\toprule
\textbf{CTR}& \textbf{PPC}& \textbf{RPM}\\
\midrule
+0.295\% & +1.347\% & +0.646\% \\
\bottomrule
\end{tabular}}}
\label{online}
\end{table}






\subsection{Production A/B Test}



To evaluate \sys's effectiveness in a real production environment, we deploy \sys in the Taobao search platform to evaluate the performance gain. We use metrics of CTR, PPC, and RPM to measure \sys 's performance. 

There exist several product-retrieval channels in our search system, and we substitute one of these channels with \sys. Specifically, one of these channels deploys our baseline PinSage, we only substitute this channel's model with \sys and keep other channels unchanged. By using a standard A/B testing configuration, we conduct the comparison on 4\% of the search request traffic of the Taobao app. 
The online results for the overall traffic are shown in Table~\ref{online}. We could see an overall lift in all three important online metrics. The reason behind this is that \sys understands a user's real-time preference by utilizing the ROI-based representations, which are beneficial for both user intention enhancement and personalized recommendation. 

\subsection{System Efficiency and Effectiveness}
We empirically evaluate the efficiency and effectiveness of \sys, in both online and offline settings. 

\begin{figure}[tpb!]
    \centering
    \includegraphics[width=0.6\linewidth]{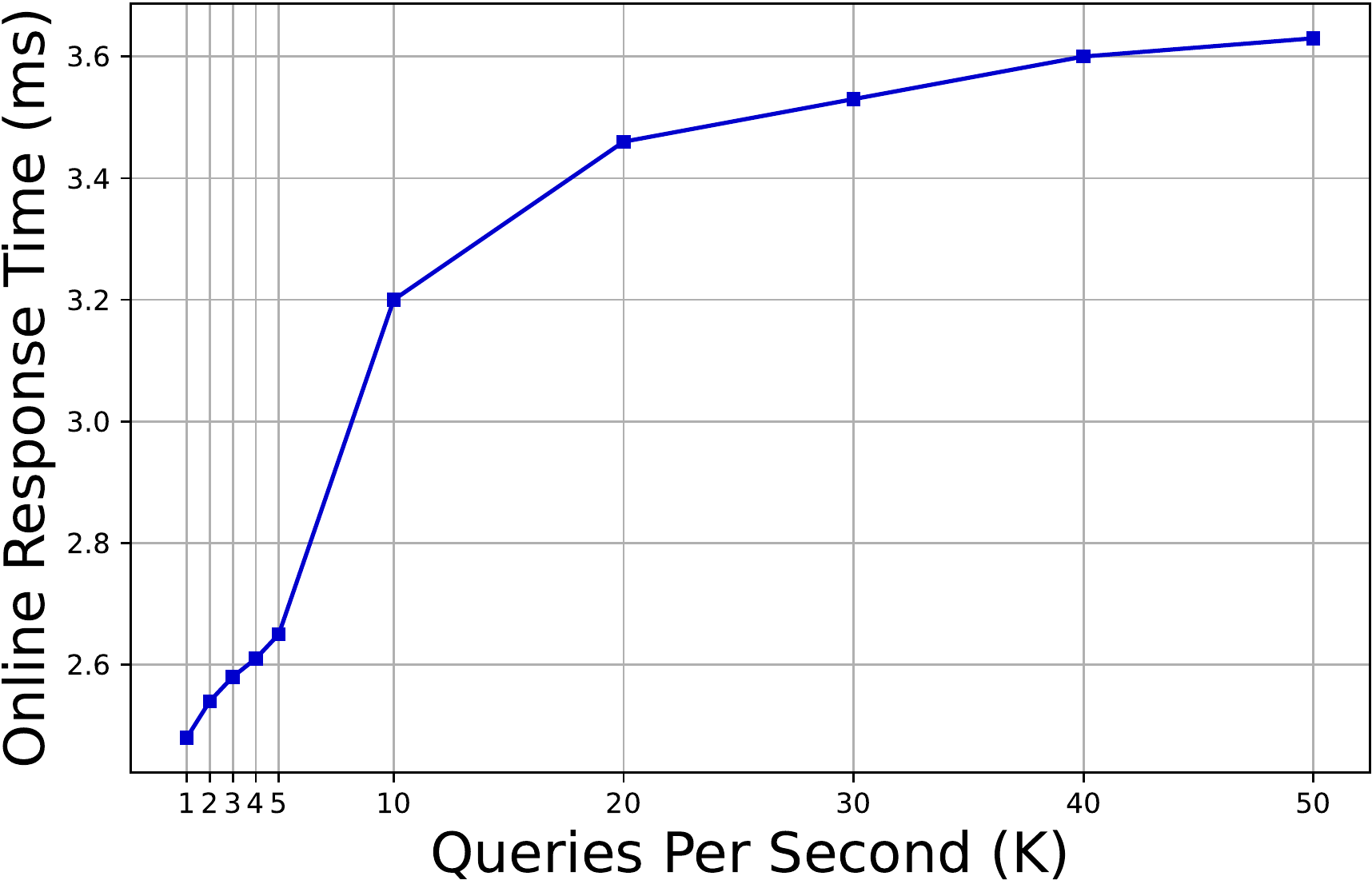}
    \caption{Online response time (ms) versus varying requests per second.}
    \label{online_serving}
\end{figure}

\para{Online Evaluation.} 
GNNs typically introduce the prohibitively high overhead of retrieving embeddings when deployed online. Long inference time will take up too much computing resource and cause the instability of the online inference engine. Besides, if the inference time takes too long, it will greatly damage the user experience. Therefore, a recommendation system with a low inference delay will greatly increase the return on investment and increase the model's business value. To this end, we adjust a few parts in \sys 's offline settings for the online deployment. Firstly, in the online GNN module, we deploy caches for dynamically storing $k$ last visited neighbors for each user and query nodes, thus avoiding the overhead for the aggregation operation. In our production deployment, $k$ is set to 30. Besides, the cache updating is fully asynchronous from users' timely requests. In this way, we decouple the neighbor-sampling process and the neighbor-aggregation process, and significantly reduce GNNs' heavy computation of electing satisfying k-hops neighbors. Besides, to further reduce computation cost, we only conserve the most effective attention part — edge-level attention part in the multi-level attention module. 

After the mentioned precautions, we deploy \sys for online serving. In the online serving stage, the two-layer inverted indexes are stored in igraph engine. Fig.~\ref{online_serving} illustrates the online response times (rt) for different QPS (queries per second). We could see that \sys handles each request less than 3 ms in average, which fully satisfies the timely response need for user experience in RS. Besides, when the QPS increases, the rt corresponding increase in a slow and smooth way. When QPS increases up to 10x, the rt increases less than 2x, indicating the effectiveness and the efficiency of our designed online serving architecture. 

\begin{figure}[tpb!]
    \centering
    \includegraphics[width=0.7\linewidth]{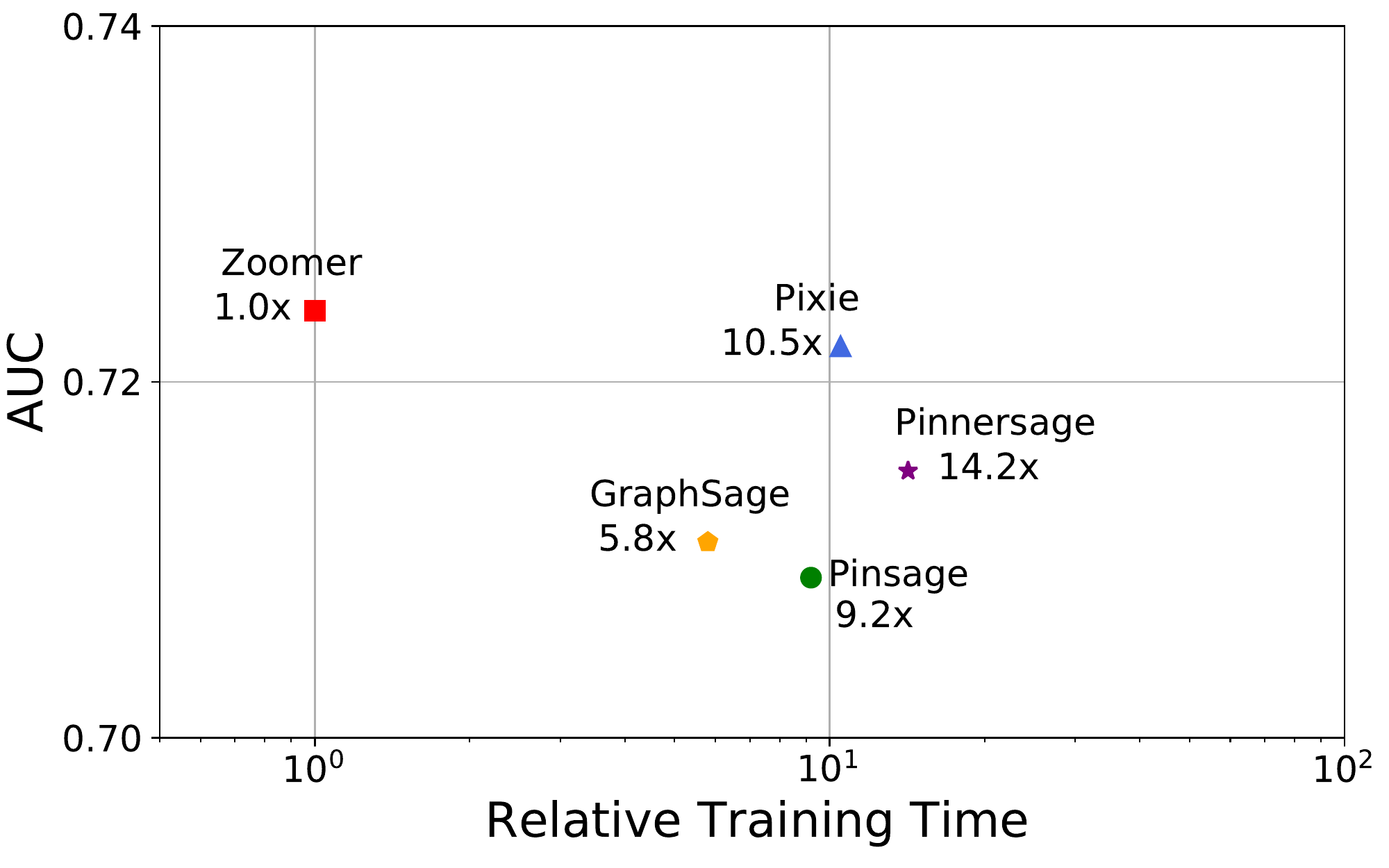}
    \caption{Efficiency versus effectiveness of \sys and baselines.}
    \label{offline_results}
\end{figure}

\para{Offline Measurement.} 
Long-term interests help search-based user behavior modeling~\cite{pi2020search}. One effective part of \sys for handling long-term interests is the focal-biased graph sampler. With the help of samplers, we can reduce the user-interaction graphs into soluble sizes. To make a fair comparison, we choose baselines with self-developed samplers, thus all methods can reduce graphs into approximate scales with importance-based samplers, and model nodes representations on reduced graphs. To this end, we choose PinSage, PinnerSage, Graphsage, and Pixie to compare. We evaluate the comparison between methods on the million-scale Taobao graph and set the number of sampling of all methods to be 30. In this setting, every method reduces the original graph to similar sizes and performs node modeling on this scale's graphs. Based on \sys 's ability in extracting timely interests and alleviating the information overload problem, we further reduce the processed graph to one-tenth of the original scale's  ROI by focal-biased samplers. We report the result in Fig.~\ref{offline_results}. We can see \sys achieves an average of 10X speedup by a further down-scaling compared to other baselines. Besides, though with a smaller reduced graph, \sys still outperforms other baselines in AUC, demonstrating that \sys can mitigate the information overload and improve node representations' quality by specifying focal-related ROI and multi-level attention mechanisms. 

\begin{figure}[tpb!]
    \centering
    \includegraphics[width=0.7\linewidth]{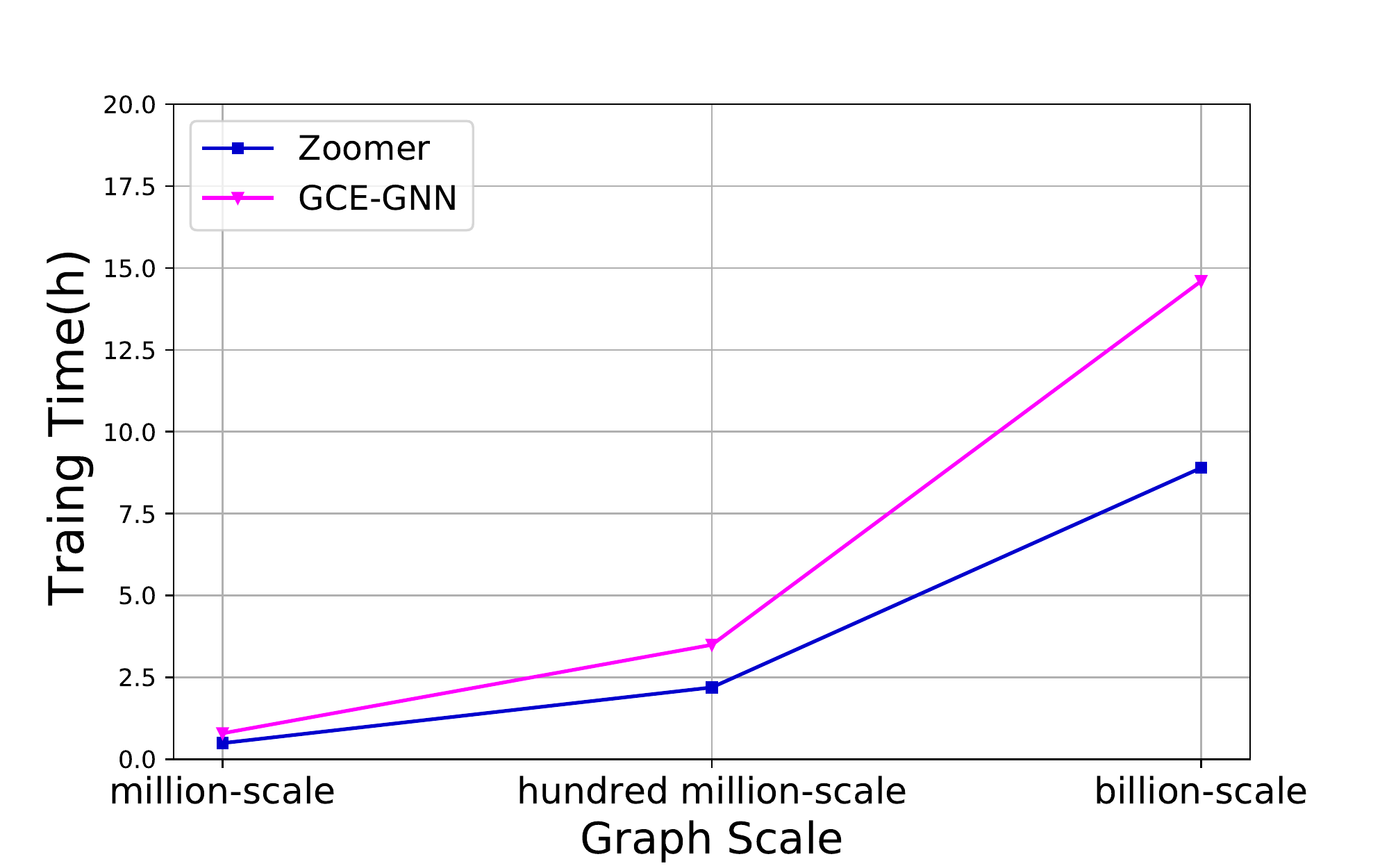}
    \caption{Training time w.r.t. graph scale.}
    \label{graph_scale}
\end{figure}

We then evaluate the scalability of \sys's distributed architecture for supporting web-scale RS service. We compare how the training time changes when the graph size increases in Fig.~\ref{graph_scale}. Specifically, we set achieving AUC equals 0.6 as a goal, and record the time cost on different graphs separately. We specify the graph sampling number to be 5 on all these graphs and perform a 2-layer \sys's multi-level attention on the sampled graphs. It could be observed that \sys is capable of handling graphs of different sizes, even if the graph size grows into a billion-scale, which demonstrates our system design is robust enough to support the web-scale industry graph training. Besides, we could see that the training cost grows rapidly when the size of graphs increases. This is because when the graph becomes bigger, it contains more interaction information, which makes training on a limited ROI a harder training process. In addition, we find \sys achieves an ideal performance with a shorter time cost than GCE-GNN in all three graphs, especially when handling huge-scale graphs (e.g., billion scale), demonstrating \sys's priority in capturing important information with limited cost. 


\subsection{Impact of Sampling Number}

\begin{figure}[tpb!]
    \centering
    \includegraphics[width=0.7\linewidth]{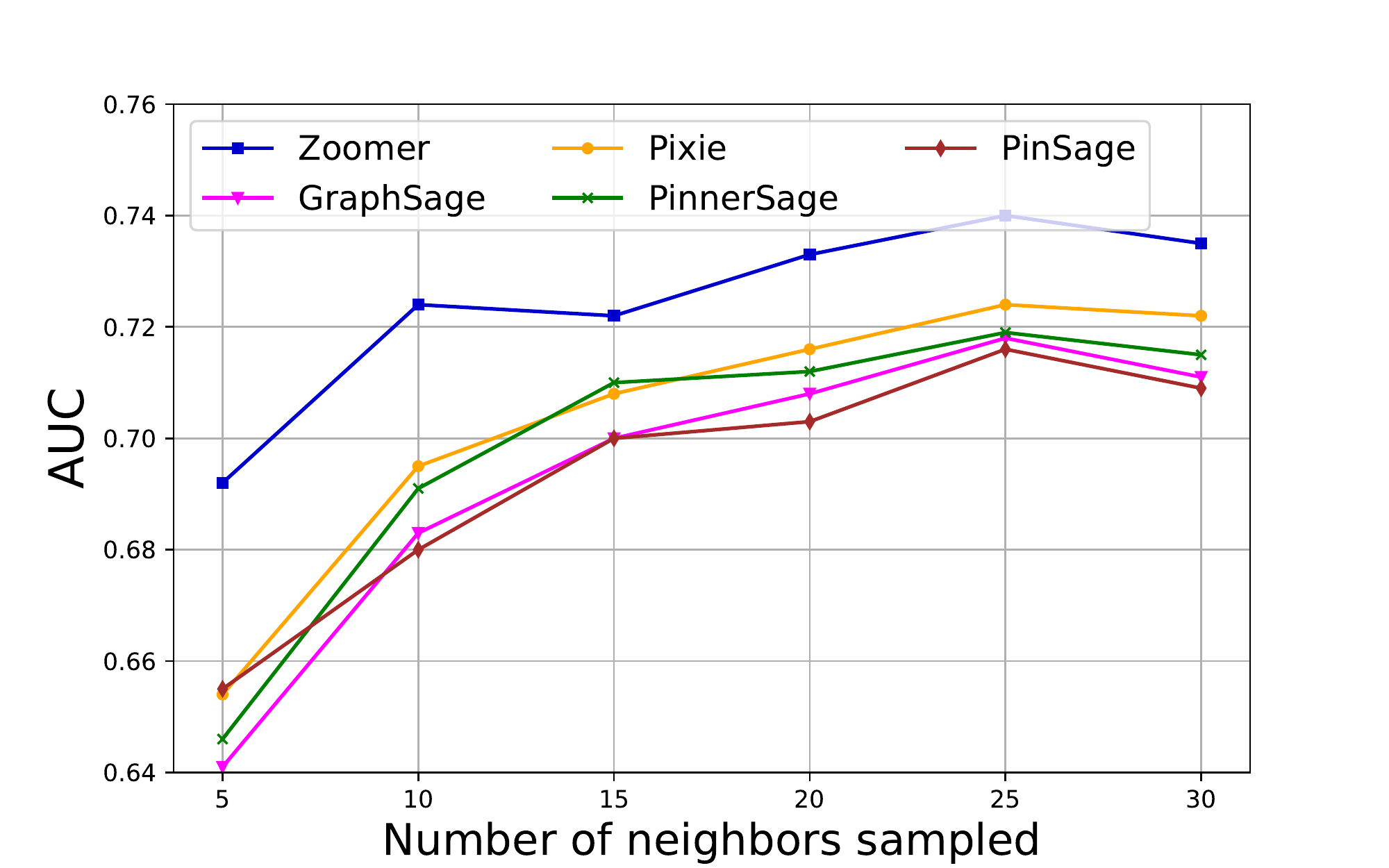}
    \caption{Effects of sampling numbers.}
    \label{randomwalk_results}
\end{figure}

The focal-oriented graph sampler is an important part of \sys, as it determines the ROI regions for performing attention strategy. To this end, we evaluate the impact of sampling neighbors' number of nodes in the sampler. For the reason that some baselines also have samplers in their design, we report the effects of sampling number on \sys and other baselines with samplers, and the result is shown in Fig. \ref{randomwalk_results}, where the x-axis corresponds to the sampling number K of each node, and the y-axis corresponds to AUC performance. We could find \sys consistently outperforms other baselines. Besides, it can be seen that \sys gets a large margin performance gain when the number being sampled is small. It indicates that \sys finds a more informative sub-graph with a limited budget, which is a supposing characteristic for handling long-term interests. Besides, some points in Fig. \ref{randomwalk_results} are contradictory with the intuition that larger graphs with more information should lead to higher model performance, as we can see that performance of all methods with K = 25 is generally better than the setting of K = 30.  This phenomenon also supports our characterization of the information overload problem, where excessive information is not necessarily beneficial for the model to learn better representations.

\begin{figure}[tpb]
\centering

 \subfigure[Fixed user.]{
  \scalebox{0.47}[0.47]{
   \includegraphics[width=1\linewidth]{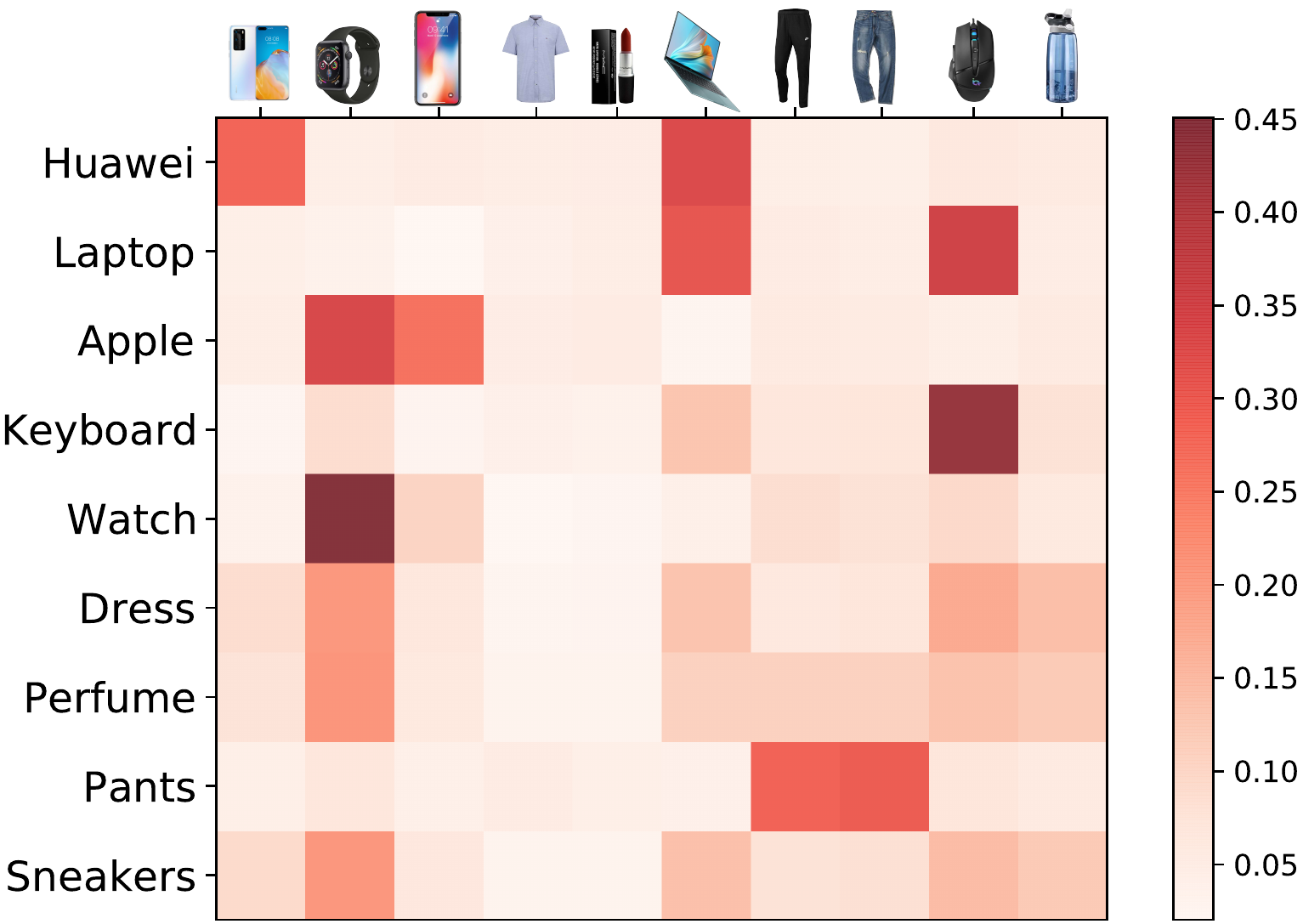}
   \label{fig:modify_q}
 }}
 \subfigure[Fixed query.]{
  \scalebox{0.42}[0.42]{
   \includegraphics[width=1\linewidth]{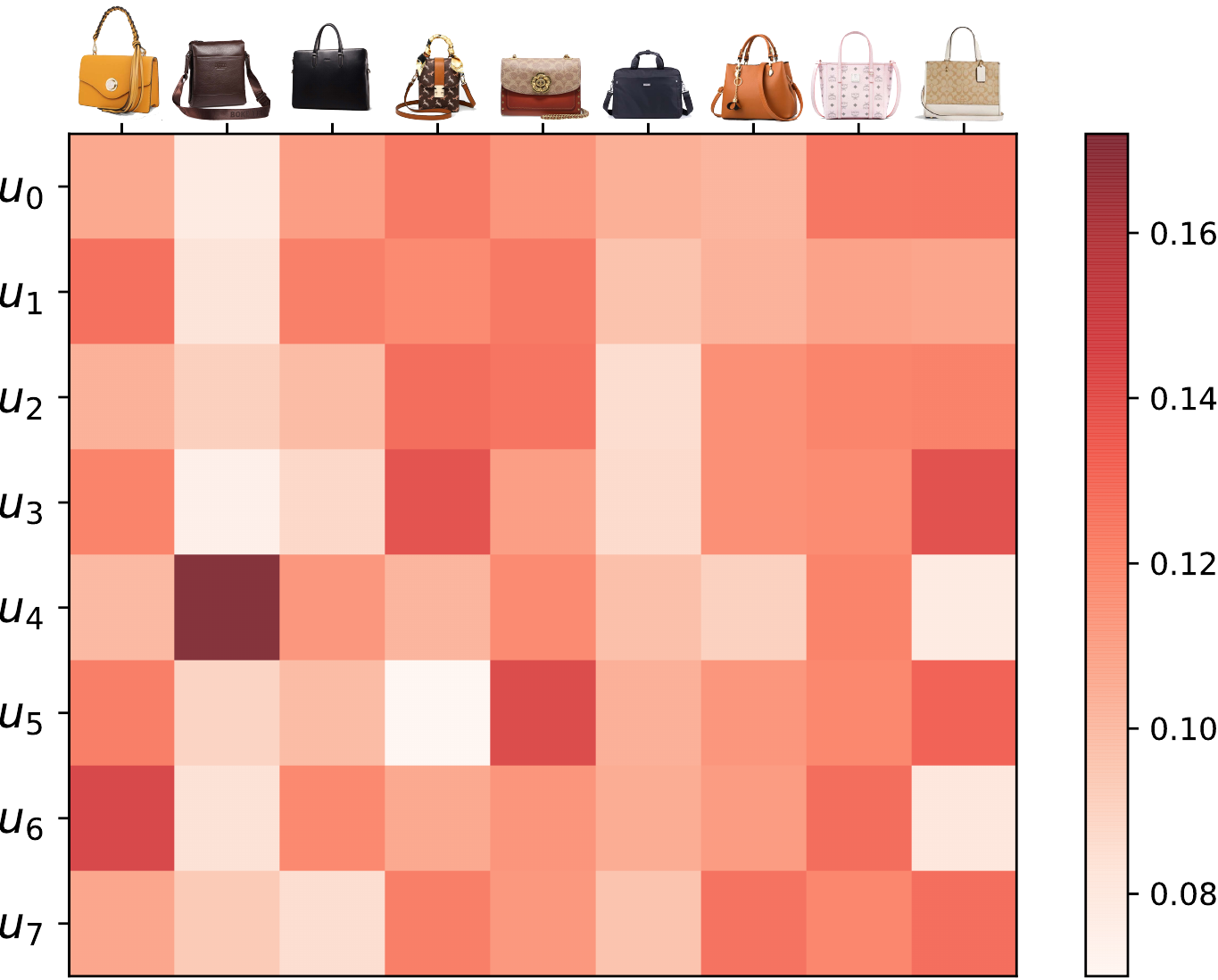}
   \label{fig:modify_u}
 }}
\caption{Heatmaps of coupling coefficients.}
\label{fig:interpretability}
\end{figure}

\subsection{Model Interpretability}
One advantage of \sys is that it can generate multi-aspect embeddings for the same ego node under different focal points. In this subsection, we visualize coupling coefficients to show that the process of ROI-based attention weights' allocation is interpretable. Fig.~\ref{fig:interpretability} illustrates coupling coefficients associated with a user node and a query node randomly selected from Taobao logs. In Fig.~\ref{fig:modify_q}, for the selected user A, each row corresponds to a query, for a specific row query $q_i$, we specify \{$q_i, u_a$\} as focal points, where $u_a$ is A's features. We randomly select 10 products from A's historical behavior log, and each column corresponds to one specific item. We calculate their corresponding edge-level weights based on their ROI and plot the weights in Fig.~\ref{fig:modify_q}. It shows that when we modify focal points for A, edge relations correspondingly change, proving \sys 's ability in adaptively modeling nodes relations according to different users' intentions. 

Fig.~\ref{fig:modify_u} illustrates coupling coefficients associated with a query node ``handbag''. We randomly select 9 item neighbors under this query in Taobao graphs. To analyze users' effects on this query's embedding in \sys, we randomly select 8 users from Taobao daily active users. In Fig.~\ref{fig:modify_u}, each row corresponds to a user and each column corresponds to an item. We could see that weights of item neighbors alter when focal points shift, thus we could assign multiple representations for node ``handbag'' under different focal points, and better meet RS's criterion of personalized recommendation. 

\section{Conclusion}
In this paper, we proposed a novel graph learning system deployed at Taobao for recommendations. Unlike traditional graph methods, we observed two characteristics of graph data in Recommender Systems: First, with the web-scale data, users generate multiple feedback in RS, which poses challenges for directly applying GNNs on data on such scale. Besides, RS suffers from the information overload problem that affects the quality of recommendations. To tackle these two challenges, we introduced concepts of focal and Region Of Interest (ROI) in graphs of RS. With an optimized focal-oriented graph sampling strategy and ROI-based information extraction from massive noisy data, \sys could achieve competitive performance with speed improvements in offline experiments and online deployment. 

\section*{Acknowledgments}
We thank the anonymous reviewers for their valuable suggestions.
This work is supported by NSFC (No. 61832001, 61972004), Beijing Academy of Artificial Intelligence (BAAI) and Alibaba.

\bibliographystyle{IEEEtran}
\bibliography{main}

\end{document}